\documentclass[10pt]{iopart}
\usepackage{graphicx,color}
\usepackage{iopams}  
\expandafter\let\csname equation*\endcsname\relax
\expandafter\let\csname endequation*\endcsname\relax
\usepackage{amsmath}
\newcommand{\ii}{\mathrm{i}} 
\begin{document}

\title[Spin-orbit coupling in a hexagonal ring of pendula.]{Spin-orbit coupling in a hexagonal ring of pendula}

\author{Grazia Salerno$^{1*}$, Alice Berardo$^2$, Tomoki Ozawa$^1$, Hannah M Price$^1$, Ludovic Taxis$^2$,  Nicola M Pugno$^{2,3,4}$ and Iacopo Carusotto$^1$}

\address{$^1$ INO-CNR BEC Center and Department of Physics, Universit\`a di Trento, via Sommarive 14,  I-38123 Povo, Italy}
\address{$^2$ Laboratory of Bio-Inspired and Graphene Nanomechanics, Department of Civil, Environmental and Mechanical Engineering, University of Trento, via Mesiano 77, I-38123 Trento, Italy}
\address{$^3$ Ket Lab, Edoardo Amaldi Foundation, Italian Space Agency, via del Politecnico snc, I-00133 Roma, Italy}
\address{$^4$ School of Engineering and Materials Science, Queen Mary University of London, Mile End Road, E1 4NS London, United Kingdom}
\ead{$^*$\mailto{grazia.salerno@unitn.it}}

\begin{abstract}
We consider the mechanical motion of a system of six macroscopic pendula which are connected with springs and arranged in a hexagonal geometry.
When the springs are pre-tensioned, the coupling between neighbouring pendula along the longitudinal (L) and the transverse (T) directions are different: identifying the motion along the L and T directions as the two components of a spin-like degree of freedom, we theoretically and experimentally verify that the pre-tensioned springs result in a tunable spin-orbit coupling. We elucidate the structure of such a spin-orbit coupling in the extended two-dimensional honeycomb lattice, making connections to physics of graphene.
The experimental frequencies and the oscillation patterns of the eigenmodes for the hexagonal ring of pendula are extracted from a spectral analysis of the motion of the pendula in response to an external excitation and are found to be in good agreement with our theoretical predictions.
We anticipate that extending this classical analogue of quantum mechanical spin-orbit coupling to two-dimensional lattices will lead to exciting new topological phenomena in classical mechanics.
\end{abstract}


\section{Introduction}

The topological effects that underlie intriguing quantum mechanical phenomena, such as the quantum Hall effect, are not the prerogative of quantum mechanical systems, but have recently been observed also in classical systems governed by Newton's equations \cite{Huber}.
This has sparked particular research interest as the robust modes that are characteristic of topological systems~\cite{generaltopology} are especially promising in view of applications. 
In some types of classical mechanical structures, there can be topologically-protected zero-frequency modes~\cite{Lubensky, Sussman, Paulose}. These may find key applications in the emerging field of acoustic metamaterials used for controlled stress design, structural engineering and vibration isolation~\cite{Huber, AluR, Pugno1, Pugno2}.
Whereas in analogue quantum Hall systems, topologically-protected finite-frequency edge states could lead to the implementation of an acoustic isolator, in which sound waves propagate along the edges of a structure without penetrating into the bulk and without being backscattered by system imperfections \cite{Alu3,Yang, Peano}. 

In classical analogues of the integer quantum Hall effect, the mechanical properties of a system should be designed so as to engineer topologically non-trivial phonon bands with non-zero Chern numbers. A first step in this research direction was a theoretical proposal for coupled pendula to simulate the Peierls phase factor of a charged particle hopping in the presence of a nonzero magnetic vector potential~\cite{Salerno}. Since then, there have been many important theoretical and experimental works demonstrating how to engineer artificial magnetic fields and topological lattice models for classical systems, such as lattices of pendula~\cite{Susstrunk,Topopendula}, coupled gyroscopes~\cite{Nash, Bertoldigyro} and acoustic crystals~\cite{Alu2,Alu,He,Pal}. 

Alongside these advances in topological classical mechanics, the photonics community has also been strongly active in the theoretical study and the experimental realization of topological lattice models \cite{Wang2009, Hafezi, Rechtsman, Simon, Carusotto} and of a spin-orbit coupling for photons \cite{Kavokin, Leyder, Yin, Khanikaev, Sala_soc, Nalitov}. 
Spin-orbit coupling in such photonic systems has been predicted to induce various topological phenomena, such as topological phase transitions \cite{Nalitov, Nalitov2}. 
A detailed theoretical and experimental study of such a spin-orbit coupling for a hexagonal ring of exciton-polariton microcavities was reported in~\cite{Sala_soc}, where two polarization states provided the pseudospin degrees of freedom.

In the present work, we show that similar physics can be observed also in systems governed by Newtonian classical mechanics.
Towards this goal, we theoretically and experimentally investigate an analogous mechanical model consisting of six pendula arranged in a hexagonal ring structure and coupled by springs. 
In particular, we observe that the frequency spectrum strongly depends on the ratio between the rest length of the springs and the equilibrium distance between neighbouring pendula. 
A mismatch between these two quantities results in a finite pre-tensioning of the springs which, as first anticipated by~\cite{Kariyado}, is responsible for different effective spring constants along the longitudinal (L) and the transverse (T) directions with respect to the spring axis.
This is analogous to how the coupling between neighbouring sites depends on the two polarization states in the polariton lattice of~\cite{Sala_soc}. 
On this basis, the mechanical system can also be interpreted as being subject to an effective spin-orbit coupling, as we show in the following.

This article is organized as follows. 
In section \ref{sec:model} we introduce the mechanical system under consideration and we theoretically review the origin of the spin-orbit coupling term, first in an infinite honeycomb lattice, then in the hexagonal ring of pendula. 
For this latter case, we discuss in detail the oscillation eigenfrequencies and eigenmodes and we classify them in terms of the symmetry of the oscillation pattern. 
In section \ref{sec:experiment} we present the experimental setup and we summarize its geometrical details and physical parameters. 
Section \ref{sec:results} is dedicated to the presentation of our experimental results, where we show that a comparison with the predictions of a theory for simple pendula gives already a qualitatively good agreement in terms of both the eigenfrequencies and the symmetry of the modes. As we then discuss, this agreement becomes quantitatively excellent when we take into account, for example, the effects of the non-zero radius of the spheres making up the pendula. 
These results confirm the presence of spin-orbit coupling in a system of pendula coupled by pre-tensioned springs and show how the strength of this spin-orbit coupling can be tuned by adjusting the amount of pre-tensioning.
Conclusions and perspectives for our work are finally discussed in section \ref{sec:conclusions}.

\section{Theoretical model}
\label{sec:model}

In this section we provide a short theoretical derivation of how the pre-tensioning of the springs gives rise to a spin-orbit coupling term in the equation of motion for the coupled pendula. 
As a first step, in subsection \ref{sec:twopendula} we show how the pre-tensioning splits the transverse and longitudinal oscillation modes of a system of two pendula. 
Then in subsection \ref{sec:honey} we will proceed with a review of the theory for the infinitely extended two-dimensional honeycomb lattice of pendula. 
Finally, in subsection \ref{sec:hexagon} we will specialize the theory to a finite geometry with a benzene-like ring of six pendula as considered in the experiment.
Throughout this section we will focus on the simplified case of simple pendula consisting of a point-mass $m$ attached to a wire of length $L$, whose natural oscillation frequency is then $\omega_0=\sqrt{g/L}$. 
Extension to a slightly more sophisticated model, taking into account the non-zero radius of the spheres making up the pendula used in our experimental setup, will be discussed in section \ref{sec:experiment}.

\begin{figure}[t]
\centering
\includegraphics[width=0.4\textwidth]{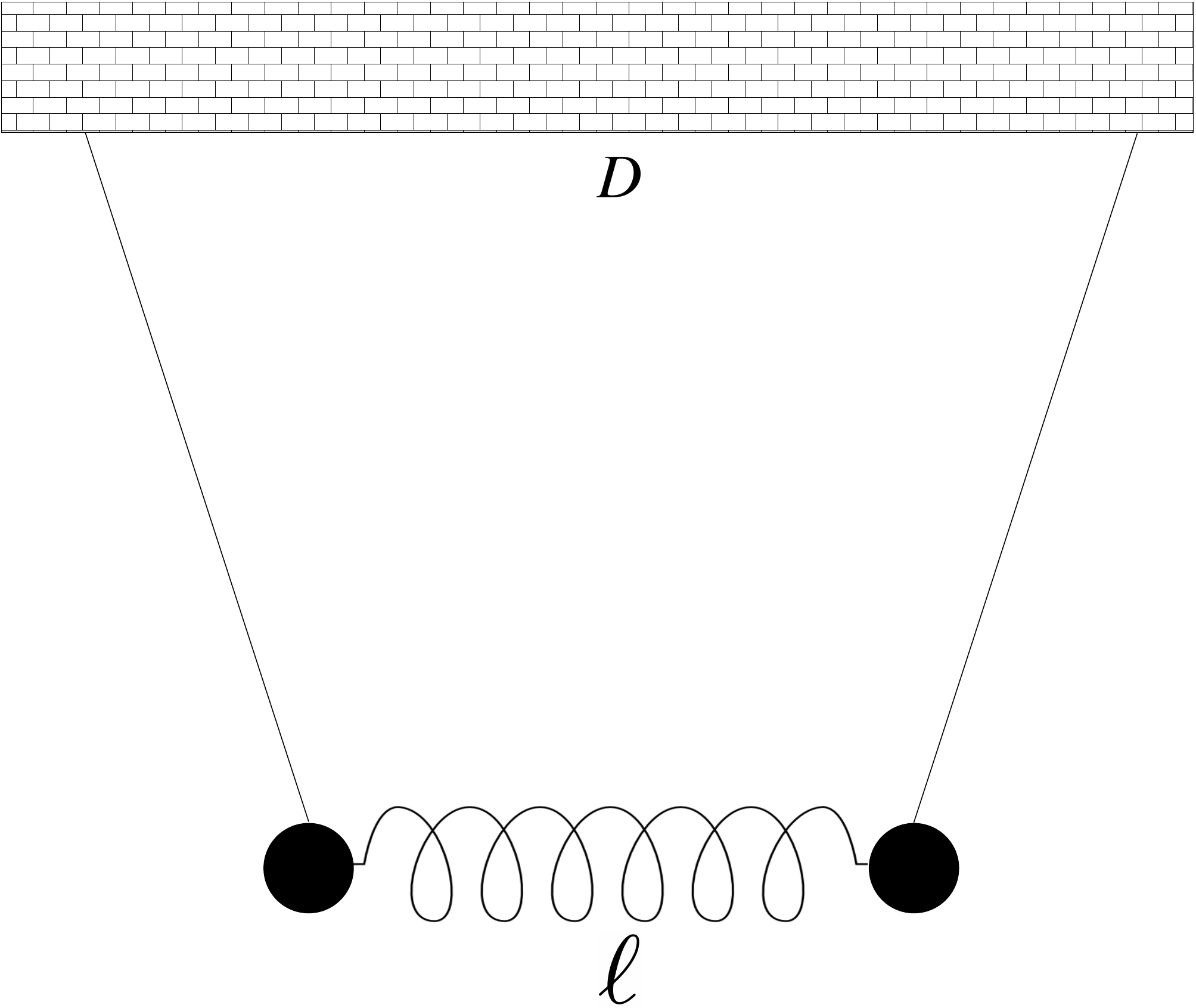}\hspace{2em}
\includegraphics[width=0.35\textwidth]{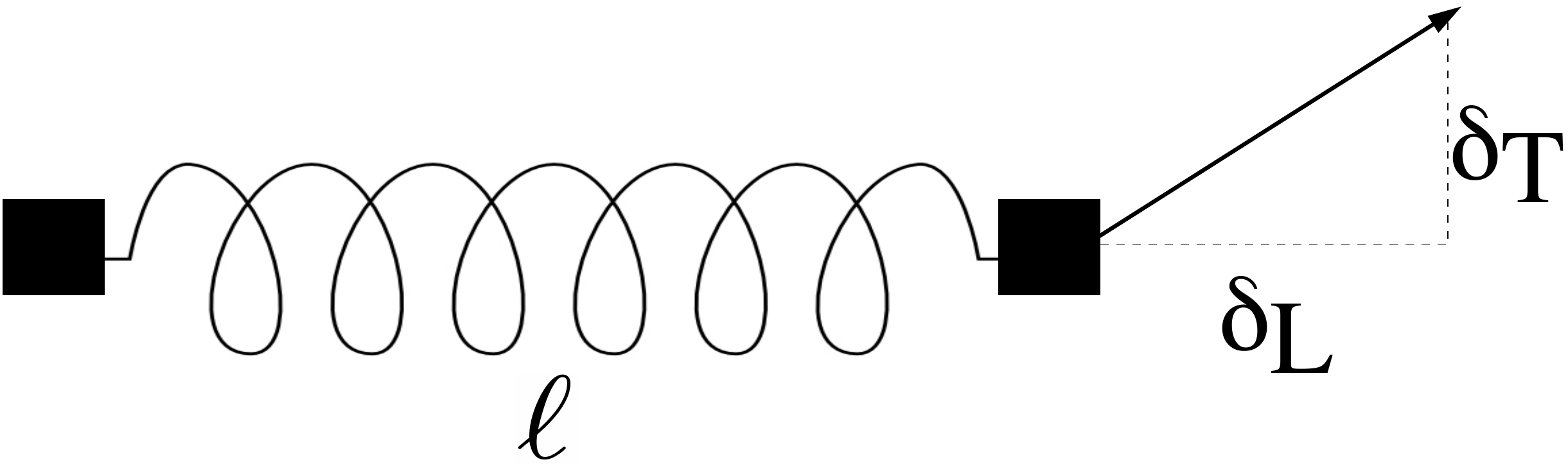}
\caption{Left panel: lateral view of a system of two coupled pendula connected by a pre-tensioned spring. 
The spring is pre-tensioned when its rest length $\ell_0$ is smaller than the distance $D$ between the hanging points of the two pendula.
As a result, in the equilibrium configuration, the total elongation $\ell$ of the spring is such that $D\geq\ell \geq\ell_0$, and the pendula have a non-zero angle with respect to the vertical direction.
Right panel: top view of the pre-tensioned spring, subjected to a further displacement that changes the relative position of the masses respectively by $\delta_L$ and $\delta_T$ along the longitudinal (L) and transverse (T) directions with respect to the link direction in the horizontal plane.}
\label{fig:spring}
\end{figure}

\subsection{System of two pendula}
\label{sec:twopendula}
We start by considering a system of two pendula coupled with a spring of spring constant $\kappa$ and rest length $\ell_0$ smaller than the distance $D$ between the hanging points, $\ell_0< D$. 
In the equilibrium configuration, the elastic force of the spring has to be balanced by the gravitational force and the tension of the wires: as a consequence, the spring is elongated to a length $\ell$ such that
\begin{equation}
D \geq\ell \geq\ell_0,
\end{equation}
so that the pendula have some non-zero angle with respect to the vertical direction, as sketched in the left part of \fref{fig:spring}. 
In the following, we refer to this feature by saying that the elongated spring in the equilibrium configuration is {\em pre-tensioned}.

As was first anticipated in \cite{Kariyado} for a system of masses and springs, such a pre-tensioned spring induces a splitting between the longitudinal (L) and transverse (T) degree of freedom of the two coupled pendula in the horizontal plane.
In fact, when one pendulum is displaced from the equilibrium position by $\delta_L$ and  $\delta_T$ along the two L and T directions, as shown in the right part of \fref{fig:spring}, the elastic energy stored in the spring grows as
\begin{equation}
U=\frac{\kappa}{2} \left(\sqrt{(\delta_L +\ell)^2+\delta_T^2}-\ell_0\right)^2.
\end{equation}
From this expression, we can see that the small oscillations along the $L$- and $T$-directions around the equilibrium position are characterized by two different effective spring constants, while the cross coupling remains zero:
\begin{equation}
\left.\frac{\partial^2 U}{\partial \delta_L^2} \right\rvert_{\substack{\delta_L=0\\ \delta_T=0 }} = \kappa \equiv \kappa_L,\qquad
\left.\frac{\partial^2 U}{\partial \delta_T^2} \right\rvert_{\substack{\delta_L=0\\ \delta_T=0} } = \kappa \left(1-\frac{\ell_0}{\ell}\right) \equiv \kappa_T, \qquad
\left. \frac{\partial^2 U}{\partial \delta_L\delta_T} \right\rvert_{\substack{\delta_L=0\\ \delta_T=0}}=0.
\label{spinorbitk}
\end{equation}

If the equilibrium length of the spring  is exactly equal to the rest length $\ell =\ell_0$, from \eref{spinorbitk} we get that the restoring force of the spring is restricted to the longitudinal direction $\kappa_L$, while the force in the transverse direction is exactly zero, $\kappa_T=0$. 
In the generic $\ell \neq\ell_0$ case, the motion along the longitudinal and transverse directions experiences different spring constants $\kappa_T \neq \kappa_L$, where the transverse one $\kappa_T$ depends on the initial elongation $\ell/\ell_0$ and becomes stronger for more pre-tensioned springs.

The small oscillations of the system of two pendula show four eigenmodes.
Their frequencies can be straightforward obtain as:
\begin{equation}
\begin{cases}
\Omega_1={\omega_0}\quad\quad\textrm{2-fold degenerate} \\
\Omega_2=\sqrt{\omega_0^2+\Omega_L^2} \\
\Omega_3=\sqrt{\omega_0^2+\Omega_T^2} 
\end{cases}
\end{equation}
with the coupling frequencies $\Omega_{L,T}$ defined as:
\begin{equation}
\Omega_{L,T} \equiv \sqrt{\kappa_{L,T}/m}.
\label{ltfrequencies}
\end{equation}
These modes include firstly a pair of eigenmodes where the two pendula oscillate in phase along either the longitudinal or the transverse direction, 
then a mode where they oscillate out of phase in the longitudinal direction and, finally, a mode where they oscillate out of phase in the transverse direction.

In the rest of this work, the oscillation of a given pendulum along the two directions will be considered as the two components of the polarization pseudo-spin where, as we show in the next subsection, the difference $\Omega_{L}-\Omega_{T}\neq 0$ provides the spin-orbit coupling in a system of many coupled pendula.

\subsection{Spin-orbit coupling in a honeycomb lattice of pendula}
\label{sec:honey}
\begin{figure}
\centering
\includegraphics[width=0.7\textwidth]{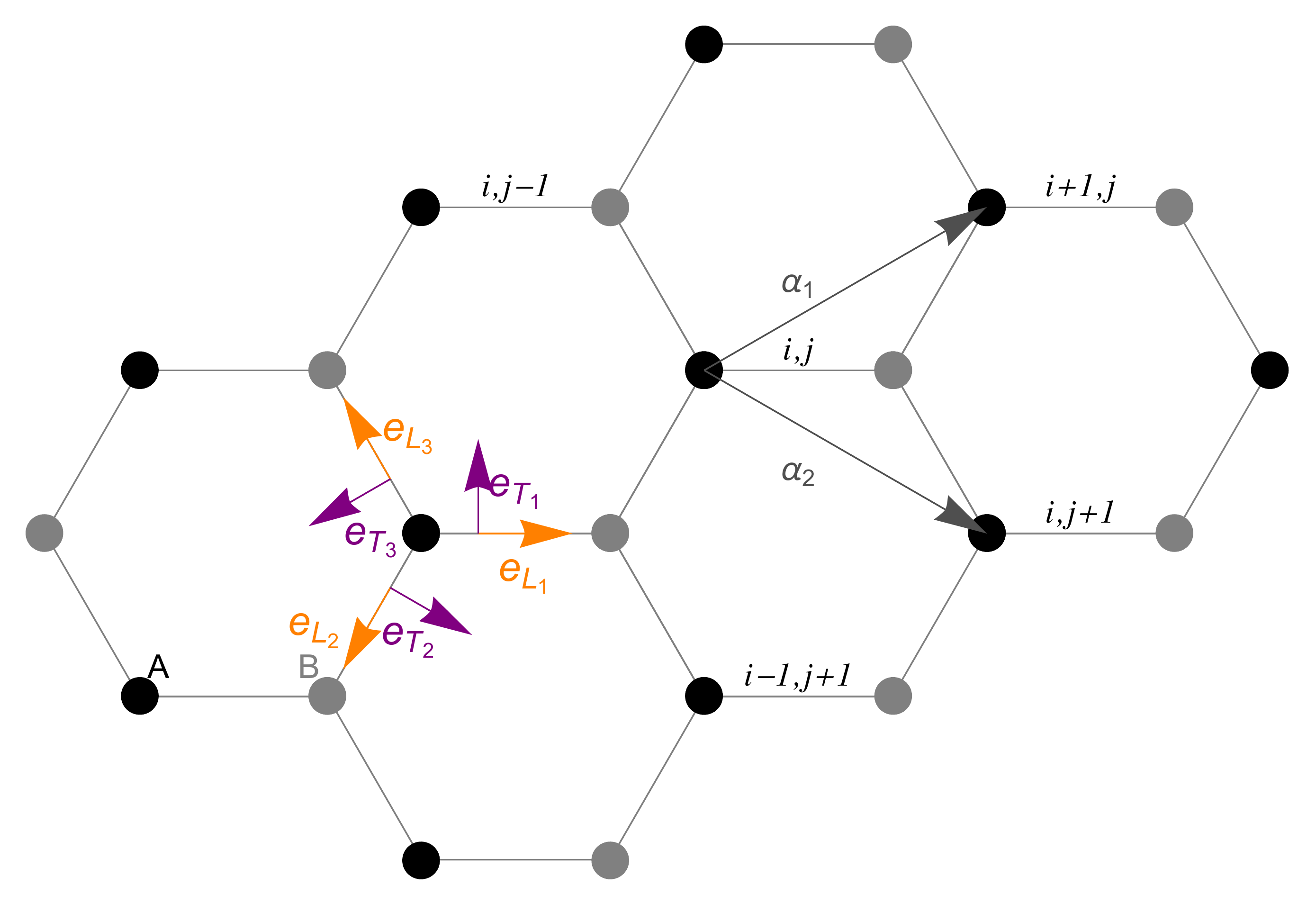}
\caption{Sketch of the honeycomb lattice, whose sites are arranged on the vertices of the hexagons.
We also show the indexing of the unit cells, each containing two lattice sites labelled as $A$ and $B$ and coloured in black and grey respectively.
The two basis vectors $\vec{\alpha}_1$ and $\vec{\alpha}_2$ that generate the lattice from the unit cell are also indicated.
For each link, orange and purple arrows represent the longitudinal $\hat{e}_{L_i}$ and transverse $\hat{e}_{T_i}$ unit vectors.}
\label{fig:vettori_SOC}
\end{figure}

The general idea of a spin-orbit coupling arising from $\Omega_L-\Omega_T\neq 0$ is best understood in the theoretically simplest case of an infinitely-extended two-dimensional honeycomb lattice of coupled pendula. 
Due to the presence of the polarization degrees of freedom, our model is similar to the $p_{x,y}$-orbital bands in a honeycomb lattice studied in~\cite{dasSarma} in the context of ultracold gases. 
Electrons in solid state graphene would instead correspond to a $p_z$-orbital band model in which there is only one valence-bond orbital per lattice site.
The honeycomb lattice is a Bravais lattice with two atoms per unit cell, labelled as $A$ and $B$ and separated by a distance $D$ equal to the lattice spacing.
The two generators of the lattice are $\vec{\alpha}_1=\left(3D/2, \sqrt{3}D/2\right)$ and $\vec{\alpha}_2=\left(3D/2, -\sqrt{3}D/2\right)$, such that the whole lattice can be recovered from one unit cell by a translation of an integer multiple of the generators. 
In this lattice, pendula are assumed to be coupled to their nearest neighbours through pre-tensioned springs such that the spring rest-length is smaller than the lattice spacing $\ell_0 < D$. 
In an infinite system, the equilibrium positions of the pendula exactly reproduce the honeycomb geometry of the hanging points, meaning that the pendula hang vertically and that the equilibrium length of the springs matches the lattice spacing $D$. 
In a more realistic finite system, this configuration can be attained by applying suitable boundary conditions at the edges of the lattice, \textit{e.g.} by keeping the position of the outermost pendula fixed.

We use the labelling shown in Fig.~\ref{fig:vettori_SOC} and introduce the following unit vectors indicated by coloured arrows in the figure:
\begin{equation}
\begin{split}
&\hat{e}_{L_1}=\left(1,0\right), \qquad \hat{e}_{L_2}=\left(-\frac{1}{2},-\frac{\sqrt{3}}{2}\right), \qquad \hat{e}_{L_3}=\left(-\frac{1}{2},\frac{\sqrt{3}}{2}\right), \\
&\hat{e}_{T_1}=\left(0,1\right), \qquad \hat{e}_{T_2}=\left(\frac{\sqrt{3}}{2},-\frac{1}{2}\right),  \qquad \hat{e}_{T_3}=\left(-\frac{\sqrt{3}}{2},-\frac{1}{2}\right).
\end{split}
\end{equation}
We denote with $\vec{a}_{i,j}=\left(a^x_{i,j},a^y_{i,j}\right)$ ($\vec{b}_{i,j}$) the displacement from equilibrium of the pendulum located on an $A$-site (a $B$-site) in the unit cell $i,j$.
Newton's equations of motion for the $A$-site pendula are the following:
\begin{equation}
\begin{split}
&\ddot{\vec{a}}_{i,j}=-\omega_0^2 \vec{a}_{i,j}\\
 &+\Omega_L^2 \Big\lbrace \left[ \left(\vec{b}_{i,j} -\vec{a}_{i,j}\right) \cdot \hat{e}_{L_1} \right] \hat{e}_{L_1} 
+ \left[ \left(\vec{b}_{i-1,j} -\vec{a}_{i,j}\right) \cdot \hat{e}_{L_2} \right] \hat{e}_{L_2} 
+ \left[ \left(\vec{b}_{i,j-1} -\vec{a}_{i,j}\right) \cdot \hat{e}_{L_3} \right] \hat{e}_{L_3} \Big\rbrace\\  
 &+\Omega_T^2 \Big\lbrace \left[ \left(\vec{b}_{i,j} -\vec{a}_{i,j}\right) \cdot \hat{e}_{T_1} \right] \hat{e}_{T_1} 
+ \left[ \left(\vec{b}_{i-1,j} -\vec{a}_{i,j}\right) \cdot \hat{e}_{T_2} \right] \hat{e}_{T_2} 
+ \left[ \left(\vec{b}_{i,j-1} -\vec{a}_{i,j}\right) \cdot \hat{e}_{T_3} \right] \hat{e}_{T_3}  \Big\rbrace,
\end{split}
\label{Newton_grapheneA}
\end{equation}
while the ones for $B$-site pendula are:
\begin{equation}
\begin{split}
&\ddot{\vec{b}}_{i,j}=-\omega_0^2 \vec{b}_{i,j}\\
 &+\Omega_L^2 \Big\lbrace \left[ \left(\vec{a}_{i,j} -\vec{b}_{i,j}\right) \cdot \hat{e}_{L_1} \right] \hat{e}_{L_1} 
+ \left[ \left(\vec{a}_{i+1,j} -\vec{b}_{i,j}\right) \cdot \hat{e}_{L_2} \right] \hat{e}_{L_2} + \left[ \left(\vec{a}_{i,j+1} -\vec{b}_{i,j}\right) \cdot  \hat{e}_{L_3} \right] \hat{e}_{L_3}
 \Big\rbrace\\  
 &+\Omega_T^2 \Big\lbrace \left[ \left(\vec{a}_{i,j} -\vec{b}_{i,j}\right) \cdot \hat{e}_{T_1} \right] \hat{e}_{T_1}
+ \left[ \left(\vec{a}_{i+1,j} -\vec{b}_{i,j}\right) \cdot \hat{e}_{T_2} \right] \hat{e}_{T_2}  + \left[ \left(\vec{a}_{i,j+1} -\vec{b}_{i,j}\right) \cdot \hat{e}_{T_3} \right] \hat{e}_{T_3}\Big\rbrace.
\end{split}
\label{Newton_grapheneB}
\end{equation}
The calculation of the normal mode dispersion is made easier by a Fourier transform to the momentum-space variables:
\begin{equation}
\begin{split}
\vec{a}_{i,j}=&\frac{1}{V}\int_{k_x \in BZ} \int_{k_y \in BZ} \begin{pmatrix}
a_x^k\\ a_y^k
\end{pmatrix} \e^{\ii \left(i \vec{k}\cdot \vec{\alpha}_1 + j \vec{k}\cdot \vec{\alpha}_2\right)} \e^{\ii \Omega_k t} \mathrm{d}k_x \mathrm{d}k_y\\
\vec{b}_{i,j}=&\frac{1}{V}\int_{k_x \in BZ} \int_{k_y \in BZ} \begin{pmatrix}
b_x^k\\ b_y^k
\end{pmatrix} \e^{\ii \left(i \vec{k}\cdot \vec{\alpha}_1 + j \vec{k}\cdot \vec{\alpha}_2\right)} \e^{\ii D \vec{k}\cdot \hat{e}_{L_1}} \e^{\ii \Omega_k t} \mathrm{d}k_x \mathrm{d}k_y.
\end{split}
\label{fouriertransform}
\end{equation}
where the integrals run over quasi-momenta in the Brillouin zone and $V$ is the total area of the Brillouin zone. 
The Brillouin zone can be taken in the form of a hexagon and is delimited by the highly-symmetric points $K$ and $K'$. 
The extra term $\e^{\ii D \vec{k} \cdot \hat{e}_{L_1}}$ accounts for the intra-cell distance between $A$ and $B$ sites.

Making use of \eref{fouriertransform} and separating different $k$ components, we project each of the equations in \eref{Newton_grapheneA} and \eref{Newton_grapheneB} along the $x$- and $y$- direction and write $-\Omega_k^2 \Psi^k_{xy} = \mathcal{D}_k \Psi^k_{xy}$, where $\mathcal{D}_k$ is the dynamical matrix in momentum space and $\Psi^k_{xy}=\left(a_x^k,a_y^k,b_x^k,b_y^k\right)^\top$.
The eigenvectors of the dynamical matrix correspond to the normal modes for a given momentum $k$.

The nature of the spin-orbit coupling becomes clearer in the circularly polarized $+/-$ basis, to which we can transform by means of the unitary matrix: 
\begin{equation}
M=\frac{1}{\sqrt{2}}\begin{pmatrix}
1 &\ii &0 &0 \\
1 &-\ii &0 &0 \\
0 &0 &1 &\ii \\
0 &0 &1 &-\ii 
\end{pmatrix}.
\label{transformation_circularbasis}
\end{equation}
In terms of the transformed vector
\begin{equation}
\Psi^k_{\pm}\equiv M\Psi^k_{xy}= \frac{1}{\sqrt{2}}\begin{pmatrix}
a^k_x + \ii a^k_y\\
a^k_x - \ii a^k_y\\
b^k_x + \ii b^k_y\\
b^k_x - \ii b^k_y
\end{pmatrix},
\end{equation}
the equation $-\Omega_k^2 \Psi^k_{\pm} = \tilde{\mathcal{D}}_k\Psi^k_{\pm}$ holds, where the explicit expression of the dynamical matrix is:
\begin{equation}
\tilde{\mathcal{D}}_k\equiv M \mathcal{D}_k M^\dagger=\begin{pmatrix}
\tilde{d}_0 & 0 &J\,V^*(k) &\Delta \, V_1^*(k)\\
0 & \tilde{d}_0 & \Delta \, V_2^*(k)  &J\, V^*(k)\\
J \,V(k) & \Delta \,V_2(k) & \tilde{d}_0 & 0\\  
\Delta\, V_1(k) & J\, V(k) & 0 &\tilde{d}_0
\end{pmatrix},
\label{matrix_graphene_soc_circular}
\end{equation}
and where we have introduced $J=\left(\Omega^2_L+\Omega^2_T\right)/2$, $\Delta=\left(\Omega_L^2-\Omega_T^2\right)/2$, and 
\begin{equation}
\begin{split}
\tilde{d}_0 &= -\frac{3}{2} J  - \omega_0^2,\\
V(k)     &= \e^{-\ii D \vec{k}\cdot \hat{e}_{L_1}} + \e^{-\ii D \vec{k}\cdot \hat{e}_{L_2}}+ \e^{-\ii D \vec{k}\cdot \hat{e}_{L_3}},\\
V_1(k) &= \e^{-\ii D \vec{k}\cdot \hat{e}_{L_1}} + \e^{-\ii D \vec{k}\cdot \hat{e}_{L_2}} \e^{-\ii 2\pi/3}+ \e^{-\ii D\vec{k}\cdot \hat{e}_{L_3}} \e^{\ii 2\pi/3},\\
V_2(k) &= \e^{-\ii D \vec{k}\cdot \hat{e}_{L_1}} + \e^{-\ii D \vec{k}\cdot \hat{e}_{L_2}} \e^{\ii 2\pi/3}+ \e^{-\ii D \vec{k}\cdot \hat{e}_{L_3}} \e^{-\ii 2\pi/3}.
\end{split}
\end{equation}
The matrix in \eref{matrix_graphene_soc_circular} can be expressed in terms of a spin operator acting on the pseudo-spin of the sublattice $A,B$ and another spin operator acting on the polarization degree of freedom. 
In our formalism, the ones acting on the sublattice degree of freedom read:
\begin{equation}
\Sigma_{\pm} =\frac{\Sigma_x \pm \ii \Sigma_y}{2} \equiv \sigma_{\pm} \otimes \mathbb{I}_2,
\label{spin_sublattice}
\end{equation}
while the ones acting on the polarization degree of freedom read:
\begin{equation}
S_{\pm} =\frac{S_x \pm \ii S_y}{2}\equiv \mathbb{I}_2 \otimes \sigma_{\pm},
\label{spin_polarization}
\end{equation}
where $\mathbb{I}_n$ is the $n$-by-$n$ identity matrix and $\sigma_{\pm}$ are defined as usual $\sigma_{\pm}= (\sigma_x \pm \ii \sigma_y)/2$ from the $2$-by-$2$ Pauli matrices.
The two operators in \eref{spin_sublattice} and \eref{spin_polarization} commute with each other.

The matrix in \eref{matrix_graphene_soc_circular} can be expanded around the $K$ ($K'$) points $\vec{k} \rightarrow (q_x, q_y-\xi 4\pi/(3\sqrt{3}D))$, for $\xi=1$ ($\xi=-1$), and for $|\vec{q}| \ll 1/D$ at the first order.
After a  gauge transformation\footnote{We perform this transformation in order to recover the Dirac-like Hamiltonian in the second term of \eref{SOCham}.} $\Psi^k_{\pm} \to \left[\begin{pmatrix} \ii & 0 \\ 0 &1 \end{pmatrix} \otimes \mathbb{I}_2 \right]\Psi^k_{\pm}$, 
we have:
\begin{equation}
\begin{split}
\mathcal{D}'_k=& \left( -\frac{3}{2} J  - \omega_0^2\right)\mathbb{I}_4 -\frac{3D}{2}J \left(\xi \Sigma_x q_x + \Sigma_y q_y\right)  
+\frac{3\Delta }{2} \left( \Sigma_x S_y - \xi \Sigma_y S_x\right)\\ & -\frac{3D\Delta}{4} \left[ S_x \left(\xi \Sigma_x q_x - \Sigma_y q_y\right) 
- S_y \left(\Sigma_y q_x + \xi\Sigma_x q_y\right)\right].
\end{split}
\label{SOCham}
\end{equation}
The first term in \eref{SOCham} is a constant and the second term is a polarization independent Dirac-like Hamiltonian, as in $p_{x,y}$-band graphene~\cite{dasSarma}; both of them are unaffected by the pre-tensioning, i.e. by $\Omega_T\neq\Omega_L$. 
More interesting are the third and fourth terms which introduce the effective spin-orbit coupling effects.
The effect of the former term is similar to that of a Rashba spin-orbit coupling, as discussed in \cite{KaneMele}, while the latter gives a trigonal warping effect \cite{Rakyta}: both of them are proportional to the $\Delta$ parameter quantifying the difference between $\Omega_{L,T}$. 

\subsection{The hexagonal ring}
\label{sec:hexagon}
\begin{figure}[t]
\centering
\includegraphics[width=0.4\textwidth]{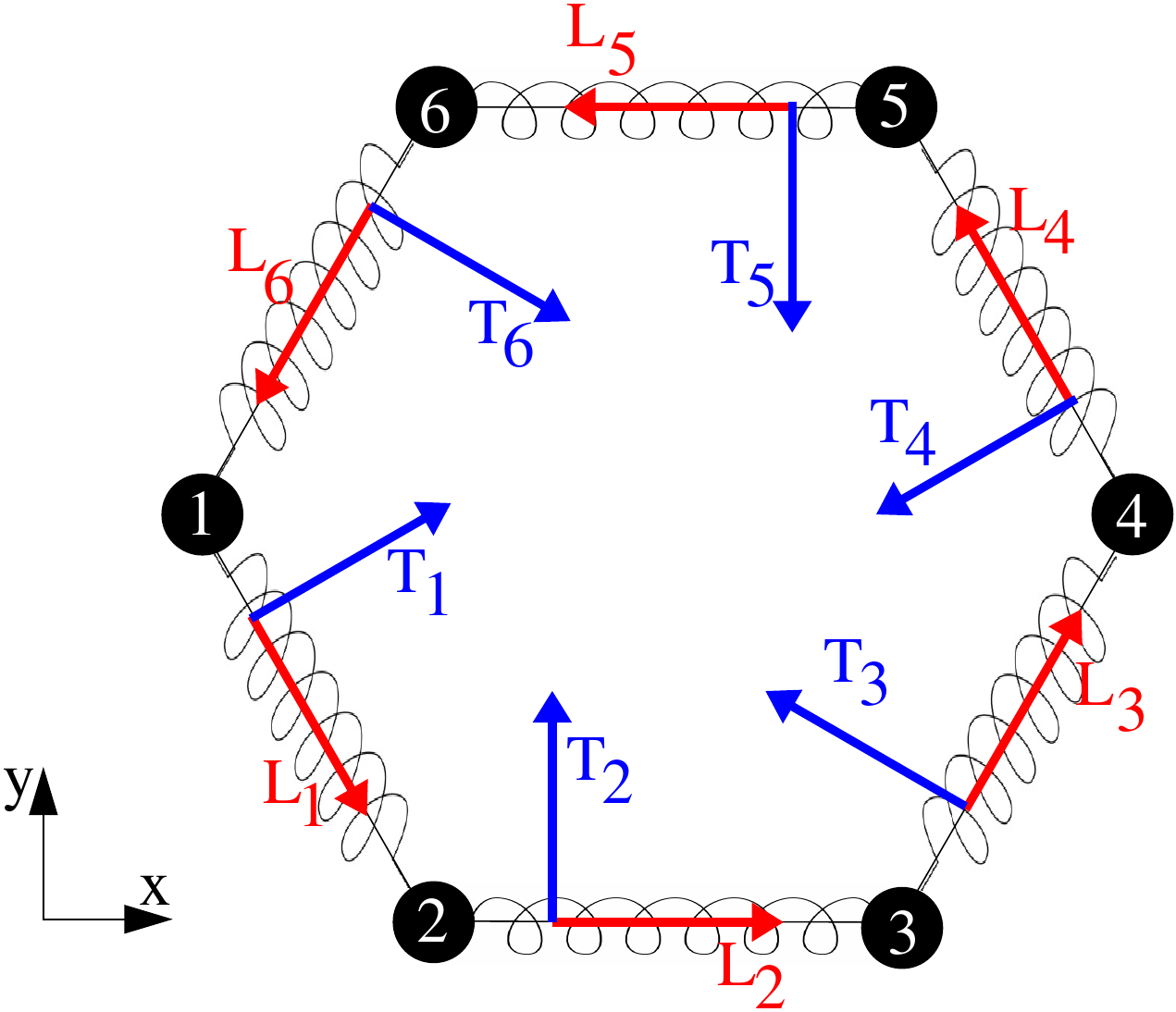}
\caption{Sketch of the mechanical system of a hexagonal ring of pendula.
The six pendula, as seen from above, are coupled with pre-tensioned springs.
The red arrows indicate the longitudinal vectors, while the blue arrows indicate the transverse ones, as used in the equation of motion in Eq.~\eref{Newton_benzene}.}
\label{fig:esagono}
\end{figure}

We conclude this theoretical section by discussing the effect of spin-orbit coupling in a benzene-like geometry consisting of a ring of six pendula arranged at the vertices of a regular hexagon. 
As in the previous sections, the springs are assumed to have a rest length $\ell_0$ shorter than the distance $D$ between the hanging points. 
In the present spatially-finite geometry, the pre-tensioning means that the equilibrium positions of the pendula lie on a hexagon of reduced side $\ell<D$, so that they make a non-zero angle with respect to the vertical direction as shown in the left panel of \fref{fig:spring} and in the left picture of \fref{fig:experimental_setup}.

\subsubsection{Equations of motion and eigenmodes}

In order to write Newton's equations of motion for the system of pendula, it is useful to separate the motion along the L and T directions and use the frequencies defined in \eref{ltfrequencies}.
For this purpose, we define unit vectors parallel and orthogonal to the direction of the links, as sketched with coloured arrows in \fref{fig:esagono}.
The explicit form of the longitudinal vectors is:
\begin{equation*}
\hat{L}_1 = \left(\frac{1}{2}, -\frac{\sqrt{3}}{2}\right),\;
\hat{L}_2 = \left(1, 0\right),\;
\hat{L}_3 = \left(\frac{1}{2}, \frac{\sqrt{3}}{2}\right),\;
\hat{L}_4=-\hat{L}_1,\;
\hat{L}_5=-\hat{L}_2,\;
\hat{L}_6=-\hat{L}_3,
\end{equation*}
while that of the transverse vectors is:
\begin{equation*}
\begin{split}
\hat{T}_1 = \left(\frac{\sqrt{3}}{2}, -\frac{1}{2}\right),\;
\hat{T}_2 &= \left(0, 1\right),\;
\hat{T}_3 = \left(-\frac{\sqrt{3}}{2}, \frac{1}{2}\right),\;
\hat{T}_4=-\hat{T}_1,\;
\hat{T}_5=-\hat{T}_2,\;
\hat{T}_6=-\hat{T}_3.
\end{split}
\end{equation*}
Expanding the elongation of each spring in this basis, Newton's equations of motion for the $i=1,\dots 6$ pendulum take the form:
\begin{equation}
\begin{split}
\ddot{\vec{\psi}}_i=-\omega_0^2 \vec{\psi}_i &+ \Omega_L^2 \left[ \left(\vec{\psi}_{i+1} -\vec{\psi}_i\right) \cdot \hat{L}_i \right] \hat{L}_i + \Omega_L^2 \left[ \left(\vec{\psi}_{i-1} -\vec{\psi}_i\right) \cdot \hat{L}_{i-1} \right] \hat{L}_{i-1} \\ 
&+ \Omega_T^2 \left[ \left(\vec{\psi}_{i+1} -\vec{\psi}_i\right) \cdot \hat{T}_i \right] \hat{T}_i + \Omega_T^2 \left[ \left(\vec{\psi}_{i-1} -\vec{\psi}_i\right) \cdot \hat{T}_{i-1} \right] \hat{T}_{i-1},
\label{Newton_benzene}
\end{split}
\end{equation}
where $\vec{\psi}_i=\left(\psi^x_i,\psi^y_i\right)$ and $\psi_i^x$, $\psi_i^y$ are the displacements of the $i$-th pendulum in the $x-y$~directions. 
Periodic boundary conditions are applied in the form $i+1\rightarrow 1$ for $i=6$ and $i-1 \rightarrow 6$ for $i=1$. 

We solve the eigenvalue problem, searching for a solution of the type $\vec{\psi}_i(t)= \vec{\psi}_i \e^{\ii \Omega t}$. 
We can cast the equation in~\eref{Newton_benzene} in a matrix form with a state vector either in the $x$, $y$ basis $\vec{\Psi}_{xy}=\left( \psi_1^x,  \psi_1^y \dots \psi_6^x,  \psi_6^y  \right)^\top$ or in the circularly polarized $+$/$-$ basis $\vec{\Psi}^{\pm}=\left( \psi_1^+,  \psi_1^- \dots \psi_6^+,  \psi_6^-  \right)^\top$, where $\psi_i^\pm=(\psi_i^x\pm\ii \psi_i^y)\sqrt{2}$.  
Regardless of the basis that is used, the system in~\eref{Newton_benzene} is:
\begin{equation}
-\Omega^2 \vec{\Psi}= \mathcal{D} \vec{\Psi},
\label{dynamicalmatrix}
\end{equation}
from which diagonalization of a $12\times12$ dynamical matrix $\mathcal{D}$ gives the frequencies of the eigenmodes.
In the general case of $\Omega_T \neq \Omega_L$, the twelve eigenmodes are grouped into a set of eight different frequencies:
\begin{equation}
\begin{cases}
\Omega_1 &= \omega_0 \quad\quad \text{2-fold degenerate} \\
\Omega_2 &= \sqrt{\omega_0^2+\Omega_L^2} \\
\Omega_3 &= \sqrt{\omega_0^2+\Omega_T^2} \\
\Omega_4 &= \sqrt{\omega_0^2+3 \Omega_L^2} \\
\Omega_5 &= \sqrt{\omega_0^2+3 \Omega_T^2} \\
\Omega_6 &= \sqrt{\omega_0^2+\frac{3}{2}\left(\Omega_L^2+\Omega_T^2\right)} \quad\quad \text{2-fold degenerate}\\
\Omega_7 &= \left(\omega_0^2+\frac{5}{4}\left( \Omega_L^2+\Omega_T^2\right)+ \frac{1}{4}\sqrt{25 \Omega_L^4 -14 \Omega_L^2 \Omega_T^2+25 \Omega_T^4}\right)^{1/2} \quad\quad \text{2-fold degenerate}\\
\Omega_8 &= \left(\omega_0^2+\frac{5}{4}\left( \Omega_L^2+\Omega_T^2\right) - \frac{1}{4}\sqrt{25 \Omega_L^4 -14 \Omega_L^2 \Omega_T^2+25 \Omega_T^4}\right)^{1/2} \quad\quad \text{2-fold degenerate}
\end{cases}
\label{eigenmodesLTcoupled}
\end{equation}

\subsubsection{Symmetry classification of the eigenmodes}

To better understand the properties of these eigenmodes, we can make use of the classification in terms of their total angular momentum first introduced for polaritons in~\cite{Sala_soc}. To see this, we consider the transformation $\tilde{T}$ which leaves the system invariant and which combines a translation $T$ that sends the $i$-th site to $i+1$-th together with a rotation $R_{\pi/3}$ of angle $\pi/3$. 
For a plane wave of wavevector $Q \in (-\pi,\pi]$ around the ring (corresponding to an orbital angular momentum $l\equiv Q/(2\pi/6) \in(-3,3]$) and uniform circular polarization $S=\pm$, we have:
\begin{equation}
\tilde{T} \vec{\Psi}^S\equiv T R_{\pi/3} \vec{\Psi}^S = \e^{\ii Q} R_{\pi/3}\vec{\Psi}^S.
\end{equation}
As the rotation operator in the $\pm$ basis acts as $R_{\pi/3} \vec{\Psi}^S = \e^{\ii S\pi/3 } \vec{\Psi}^S $, we have that:
\begin{equation}
\tilde{T} \vec{\Psi}^S= \e^{\ii \frac{\pi}{3} \left(l +S \right)}\vec{\Psi}^S, 
\end{equation}
and so we can identify $k= l+S$ as the total angular momentum. 

\begin{figure}
\includegraphics[width=0.32\textwidth]{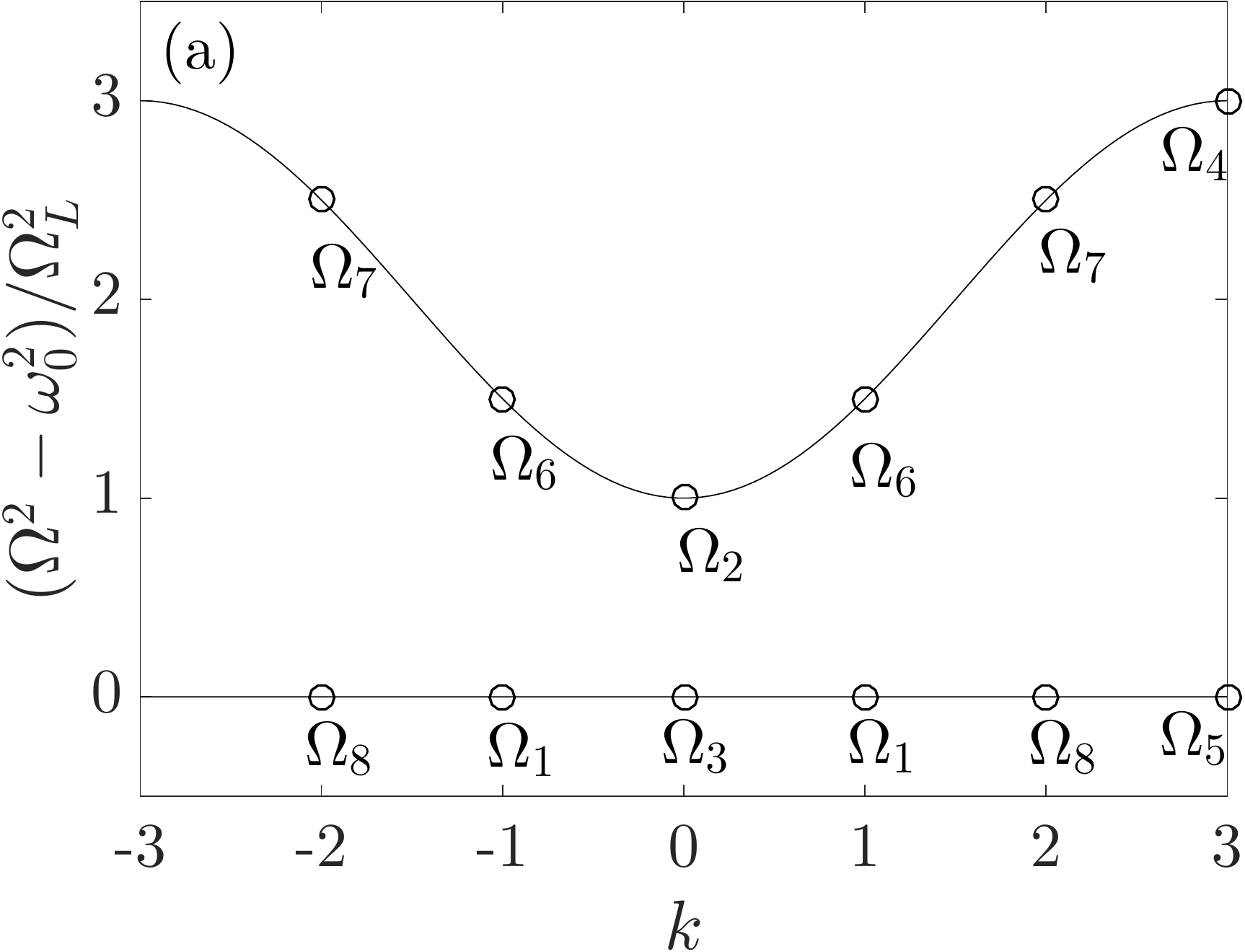}
\includegraphics[width=0.32\textwidth]{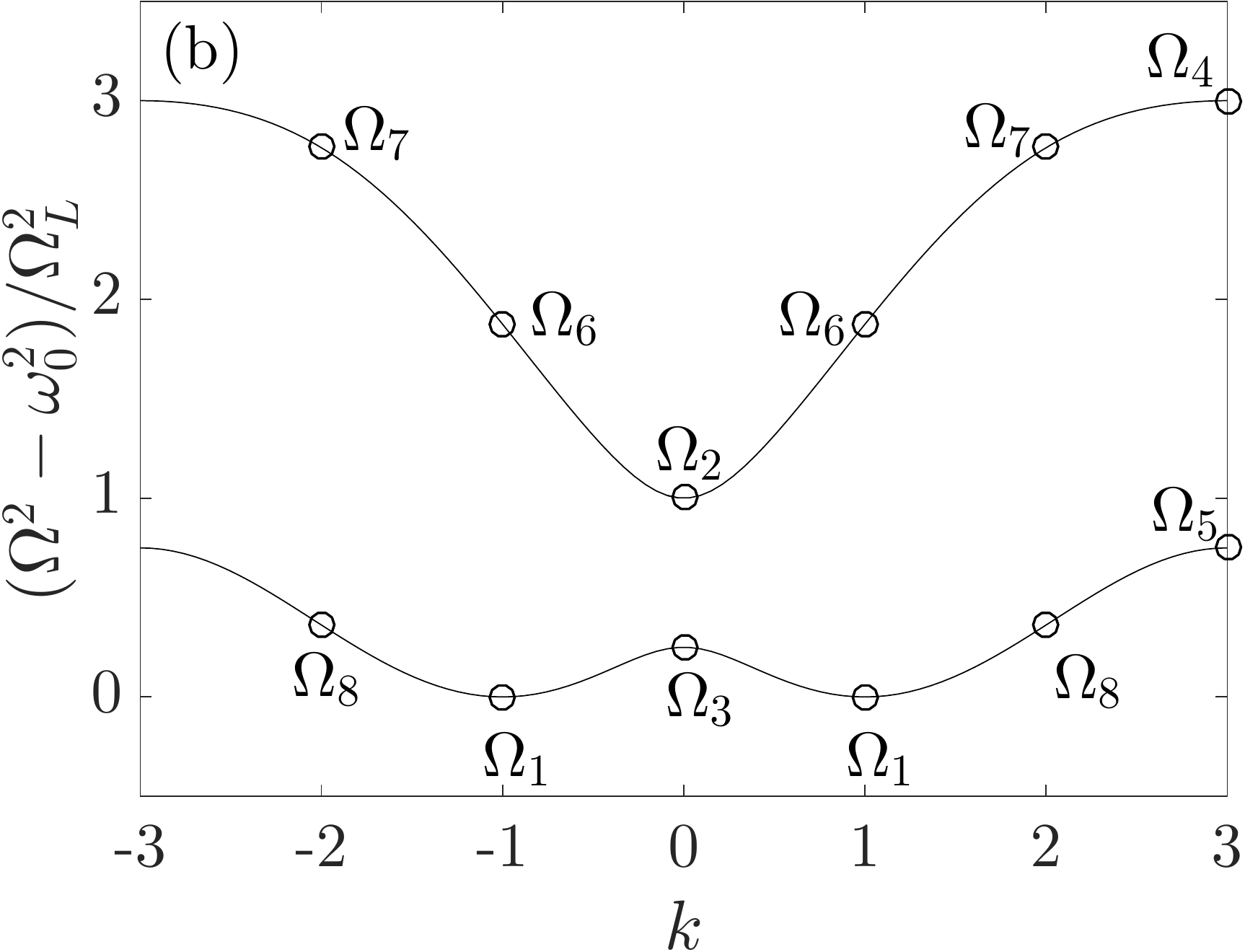}
\includegraphics[width=0.32\textwidth]{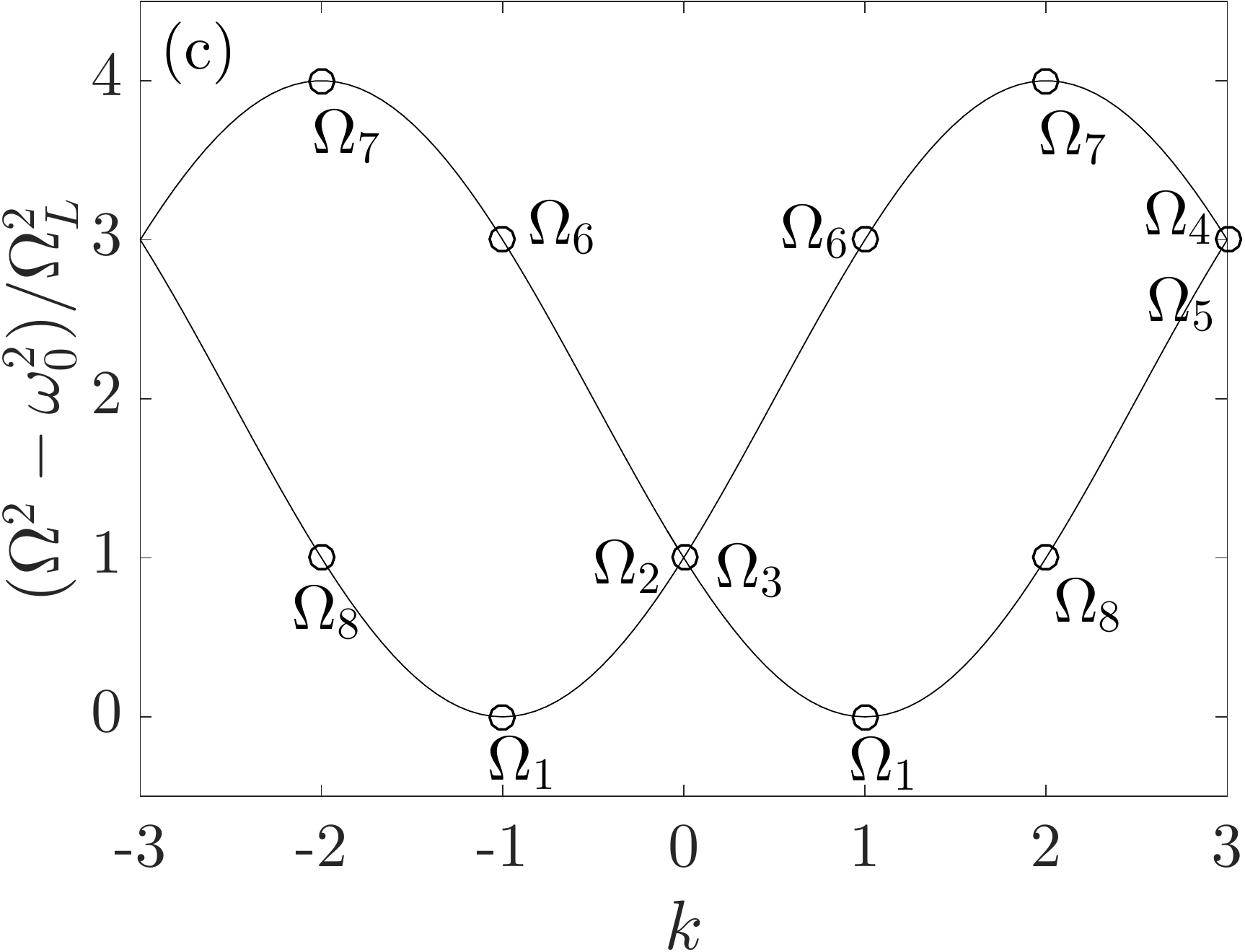}
\caption{Theoretical prediction for the normalised eigenfrequencies $\left(\Omega^2-\omega_0^2\right)/\Omega_L^2$ as a function of the total angular momentum $k$, for $\Omega_L/\omega_0=1$.
Panels a, b, c have respectively $\Omega_T/\Omega_L=0,\,0.5,\,1$.}
\label{fig:angular}
\end{figure}

\begin{figure}[p]
\centering
 \hspace{1.6cm}$\Omega_4$\includegraphics[width=0.2\textwidth]{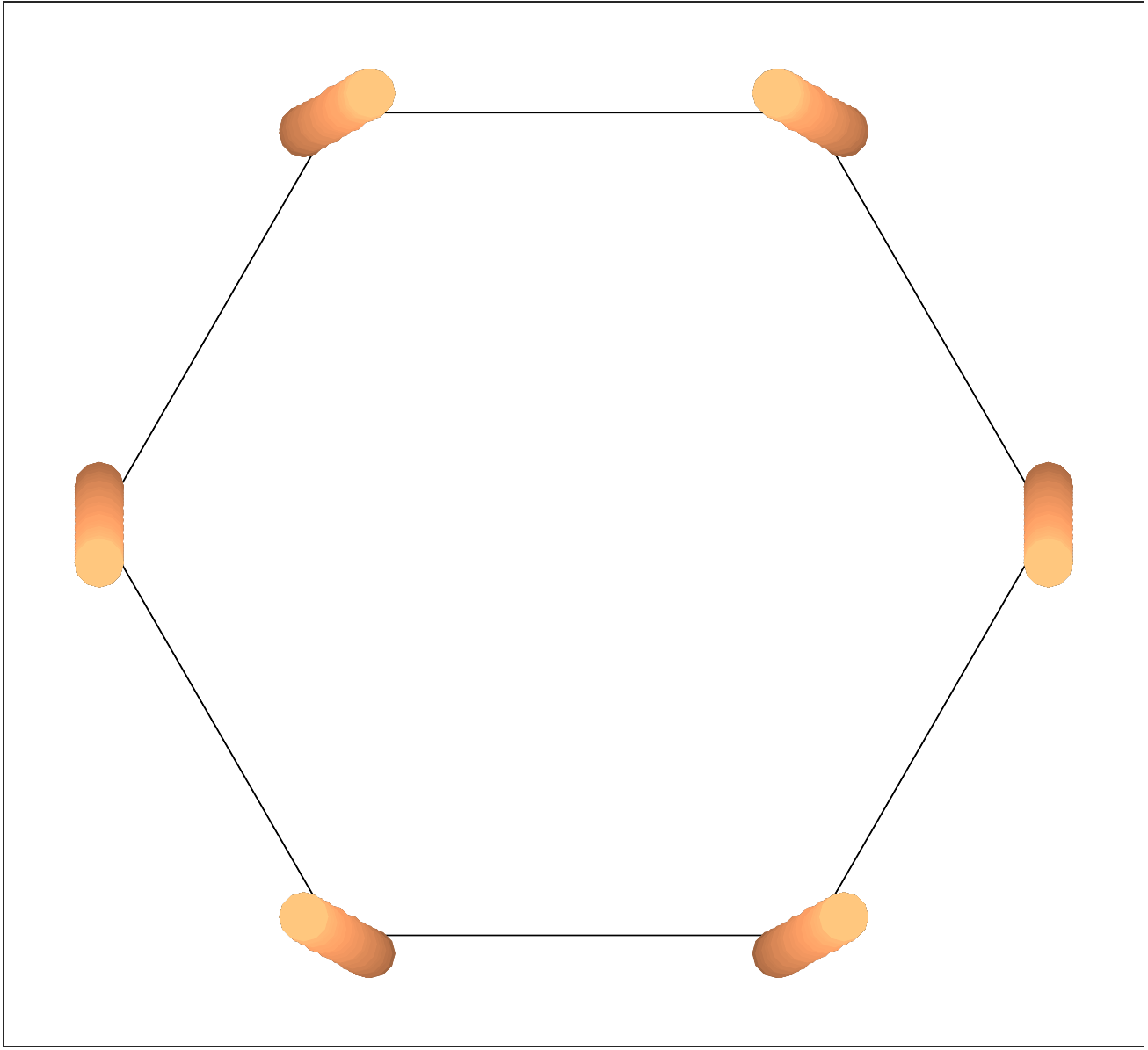} \hspace{3cm} $\Omega_4^\text{(Exp)}$\includegraphics[width=0.2\textwidth]{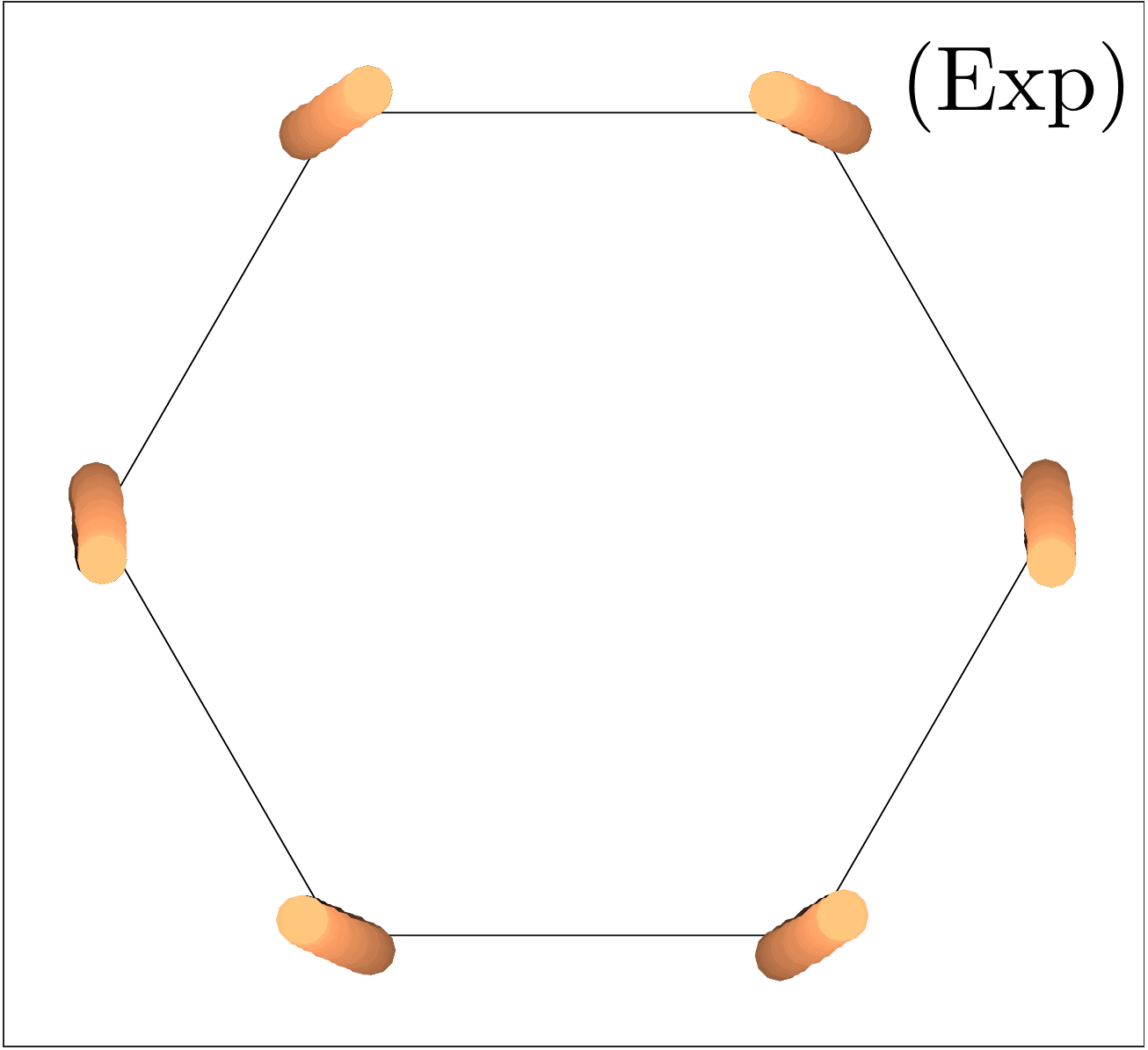}\\
$\Omega_7$ \includegraphics[width=0.2\textwidth]{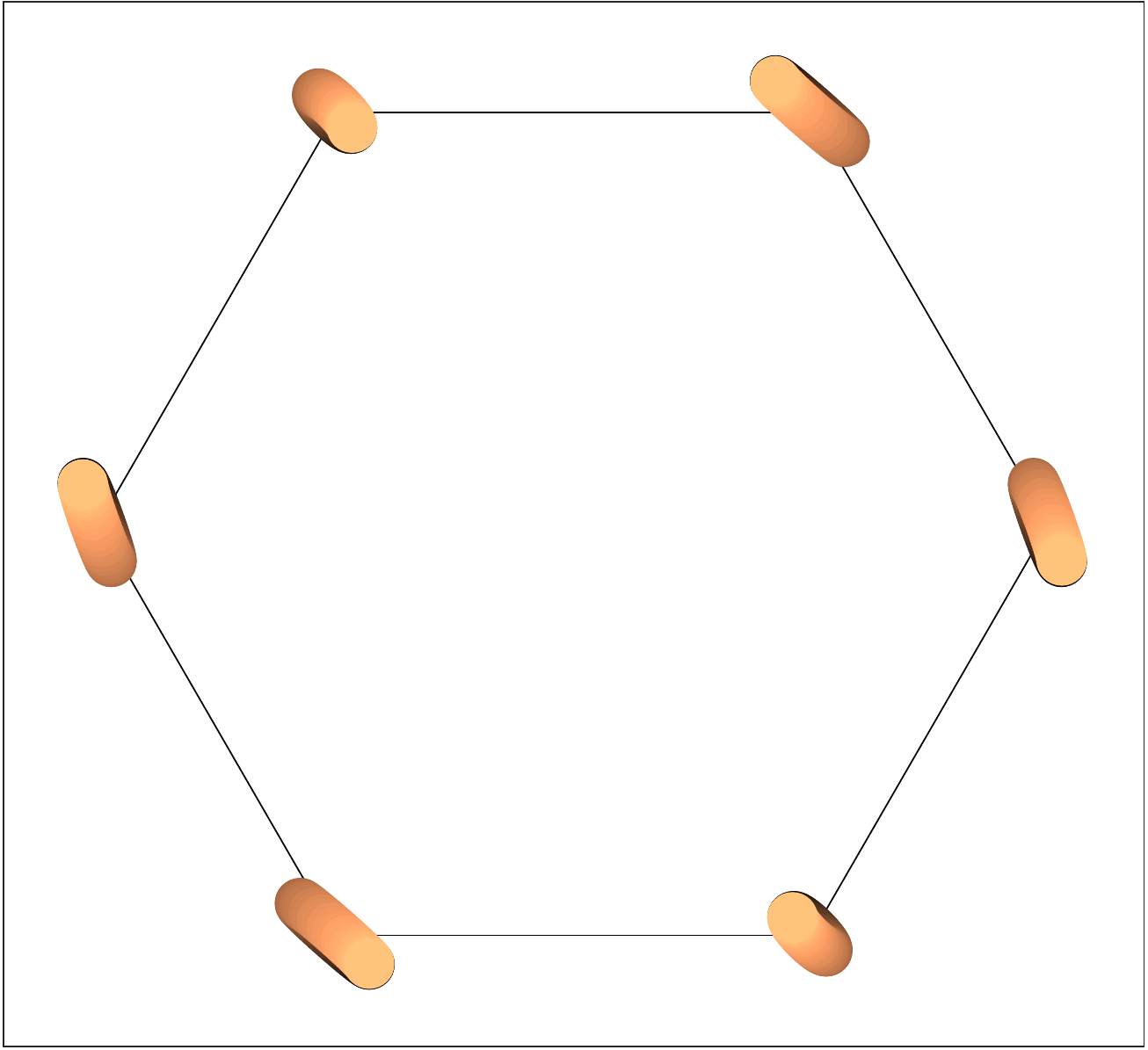}
\includegraphics[width=0.2\textwidth]{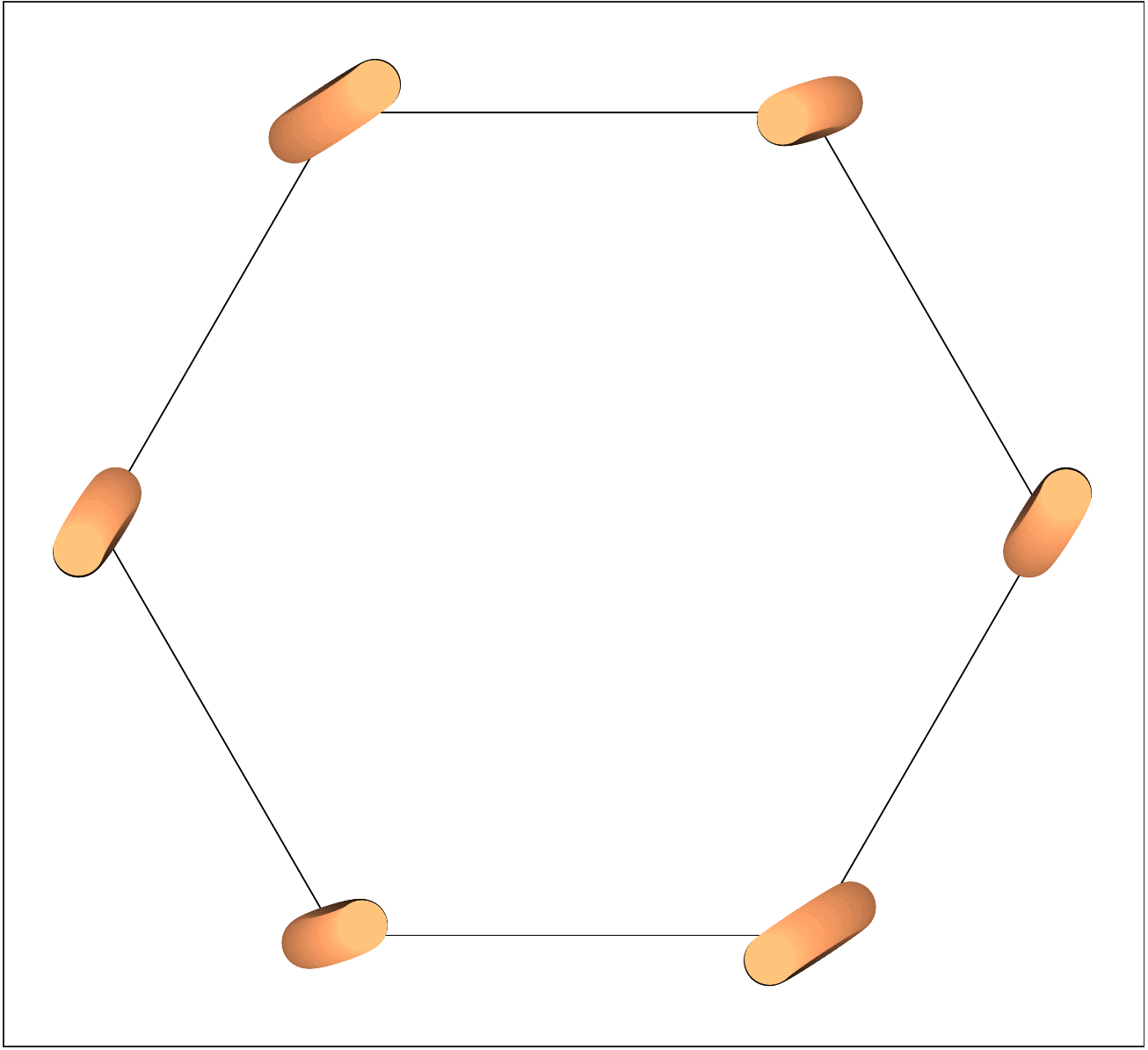} \hspace{1.8cm} $\Omega_7^\text{(Exp)}$\includegraphics[width=0.2\textwidth]{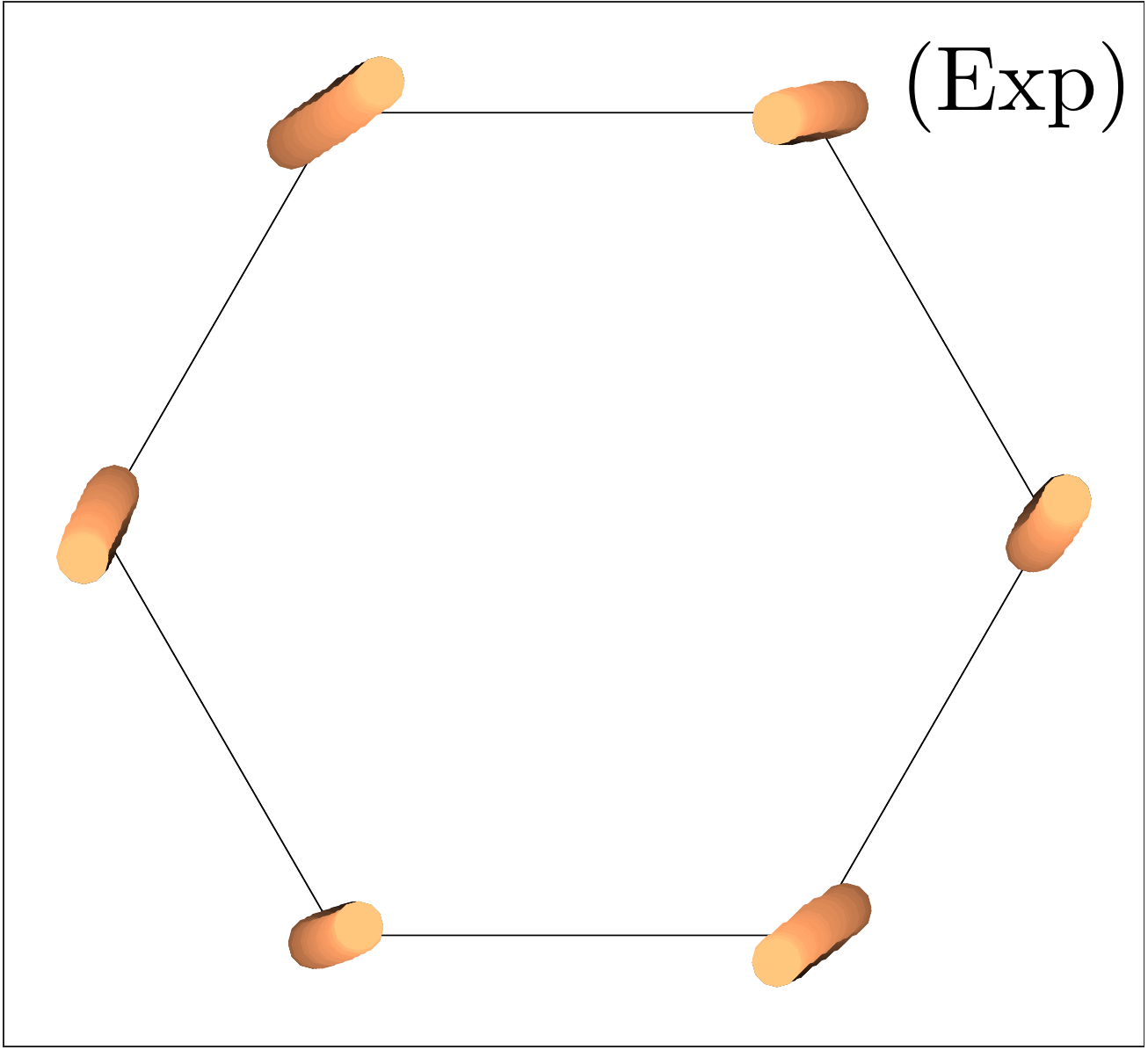} \\
$\Omega_6$ \includegraphics[width=0.2\textwidth]{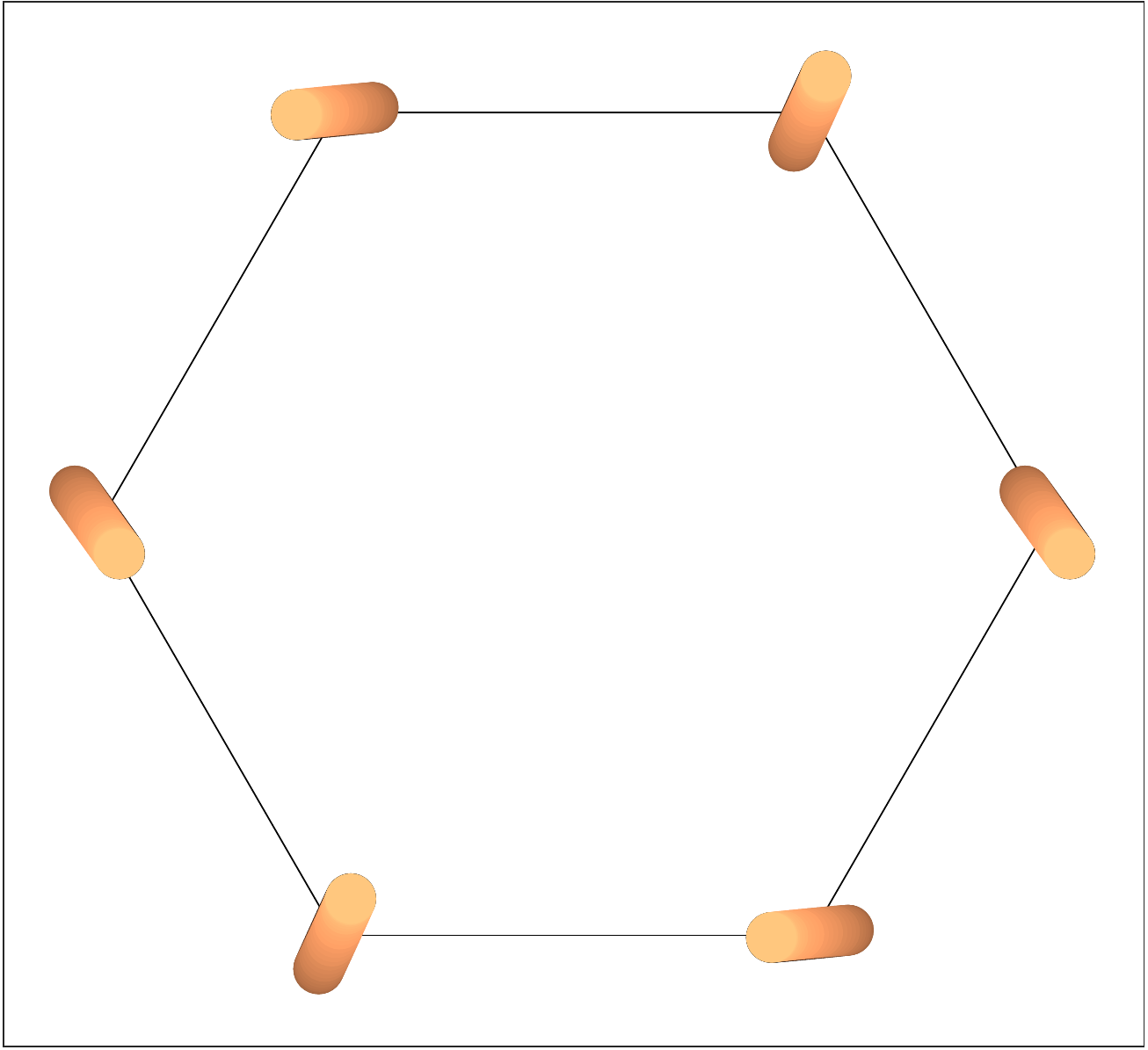}
\includegraphics[width=0.2\textwidth]{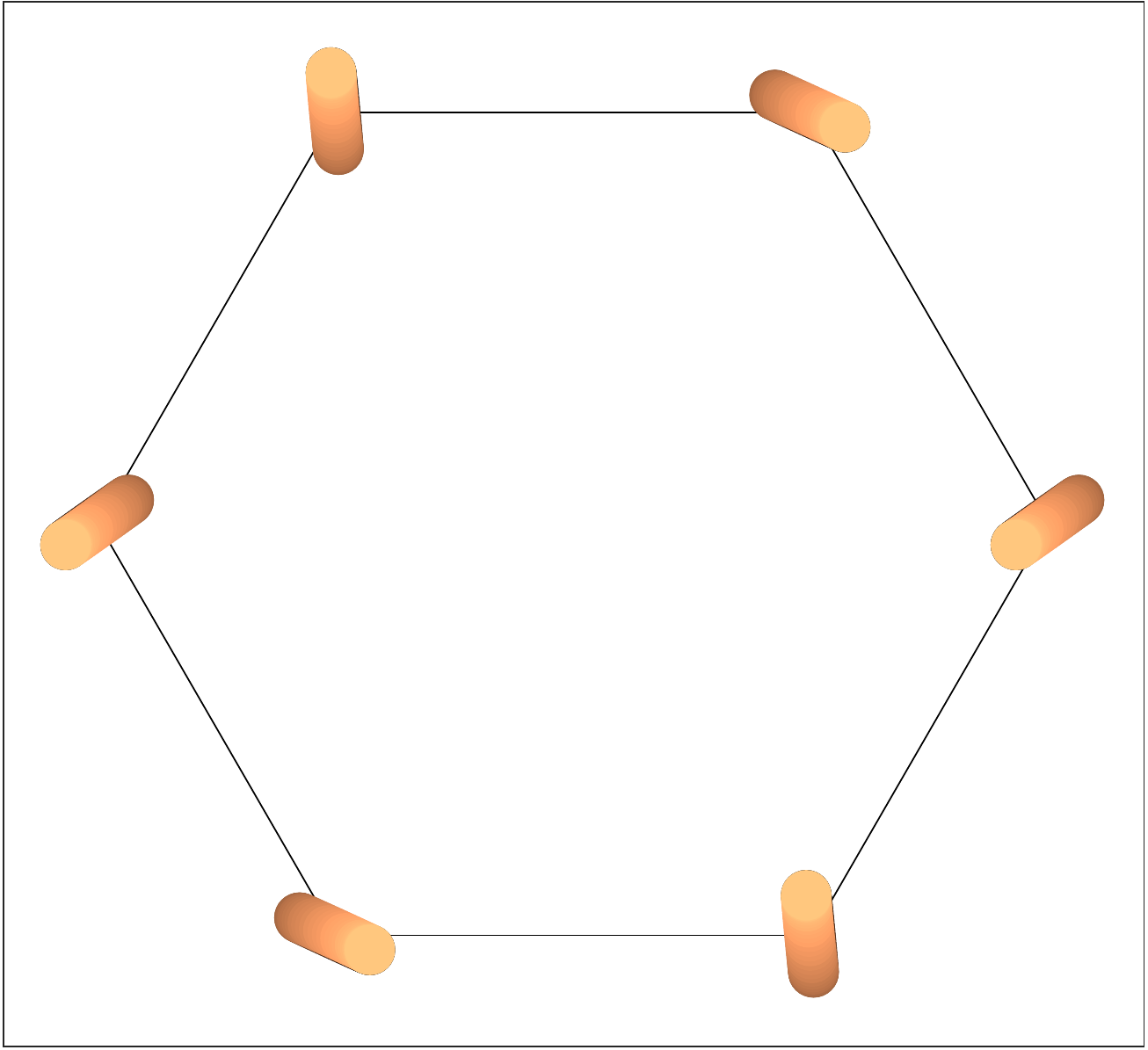} \hspace{1.8cm}  $\Omega_6^\text{(Exp)}$ \includegraphics[width=0.2\textwidth]{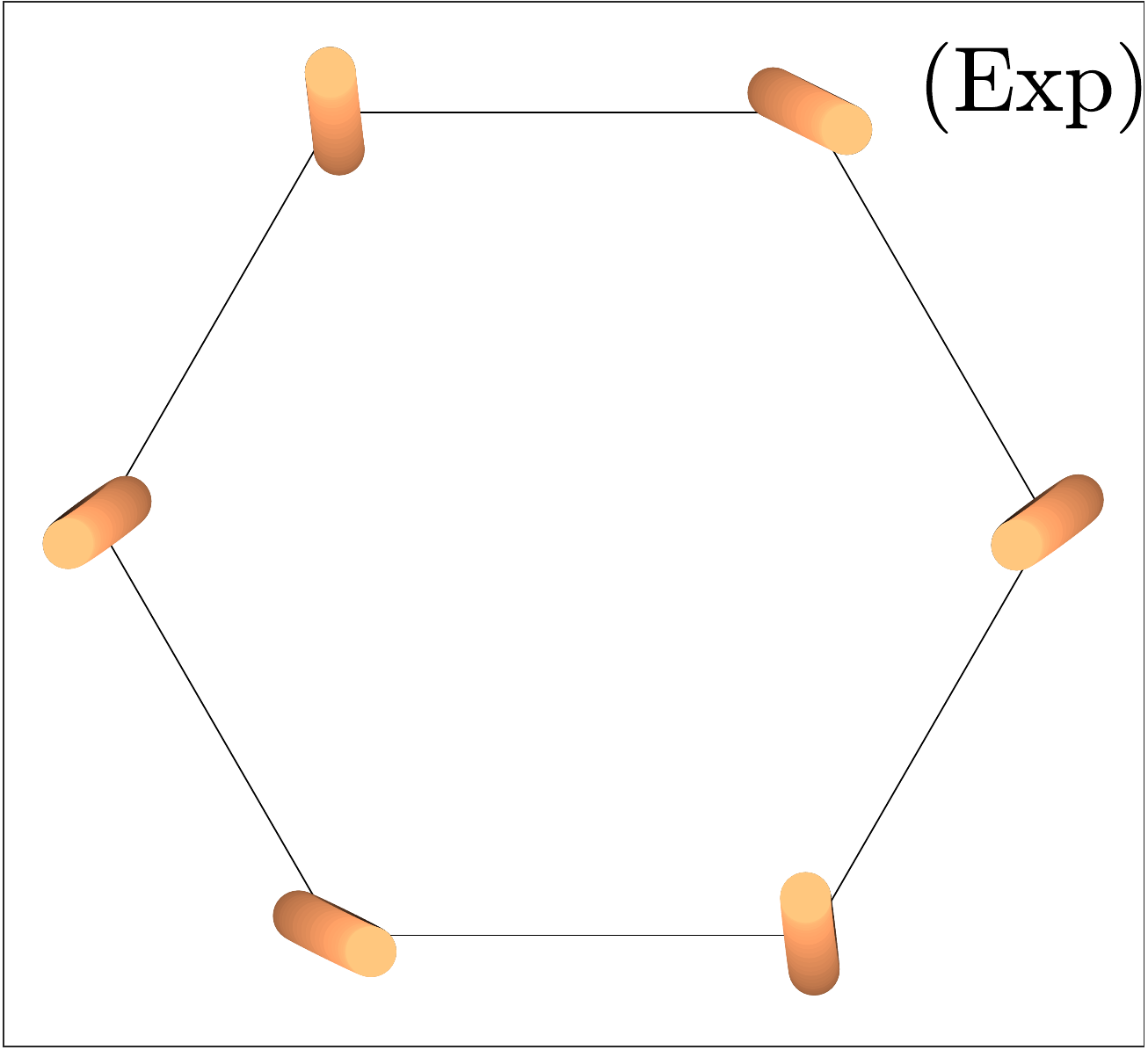}\\
\hspace{1.6cm}$\Omega_2$ \includegraphics[width=0.2\textwidth]{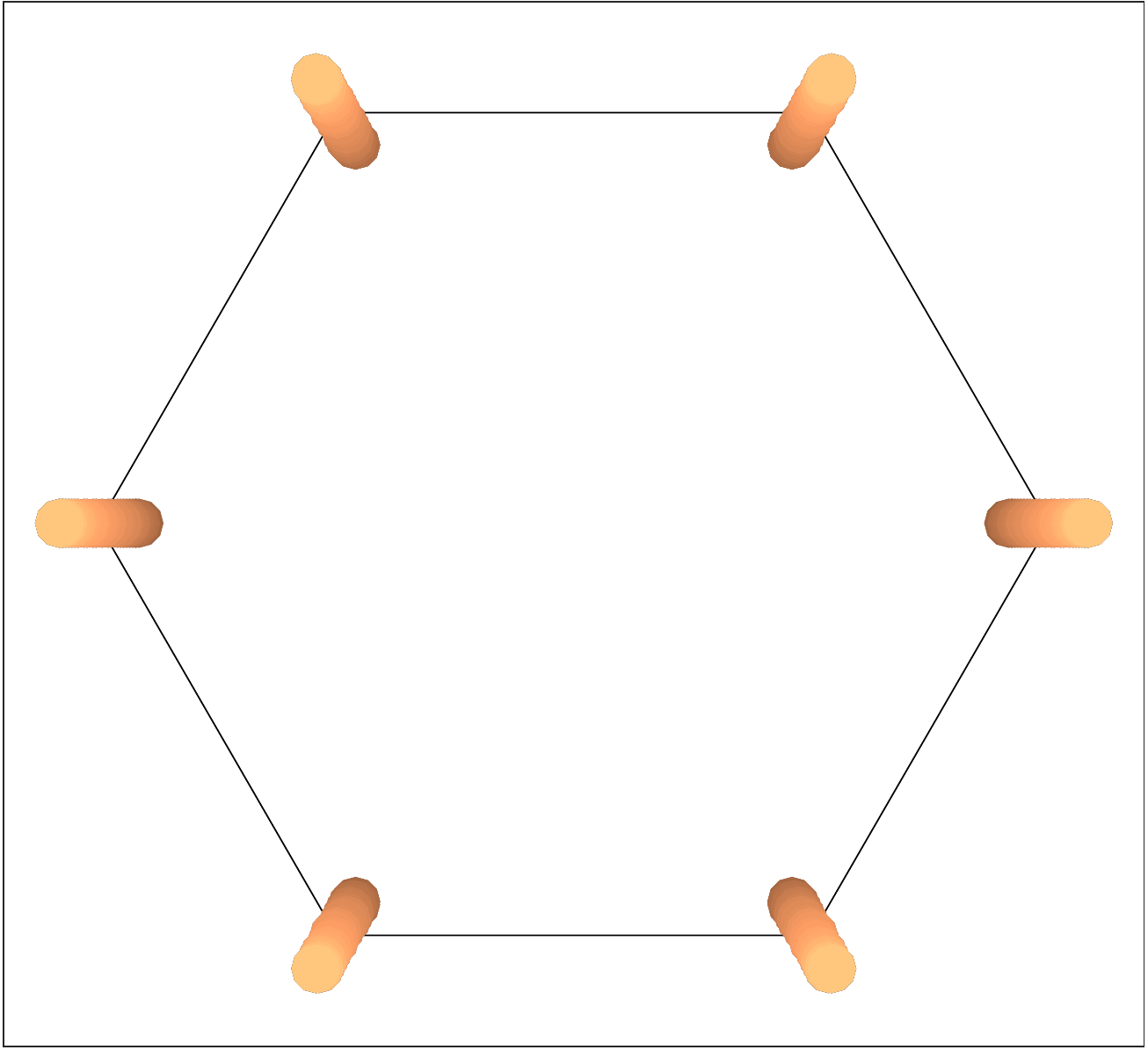} \hspace{3cm}  $\Omega_2^\text{(Exp)}$\includegraphics[width=0.2\textwidth]{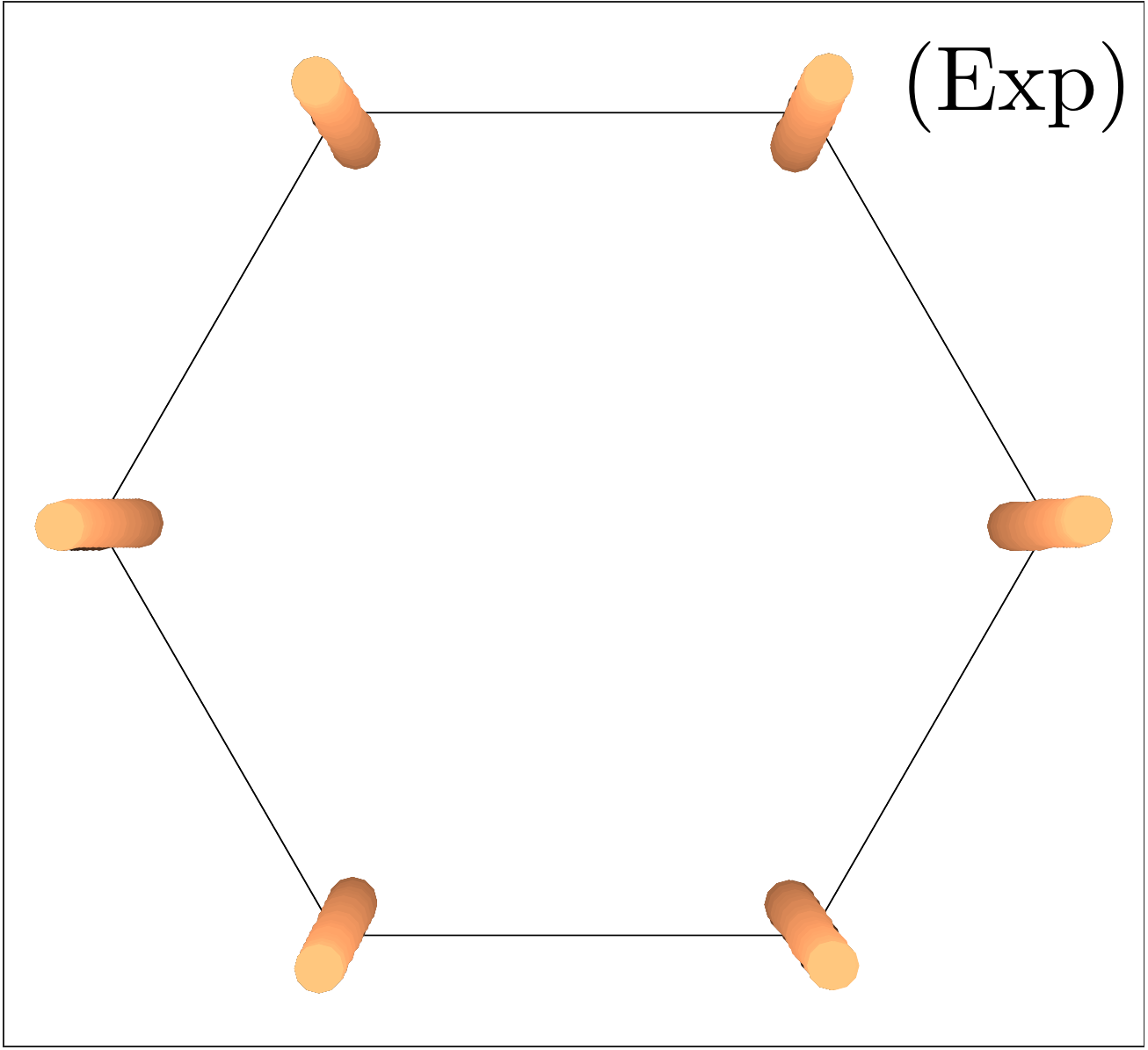}\\
\hspace{1.6cm}$\Omega_5$ \includegraphics[width=0.2\textwidth]{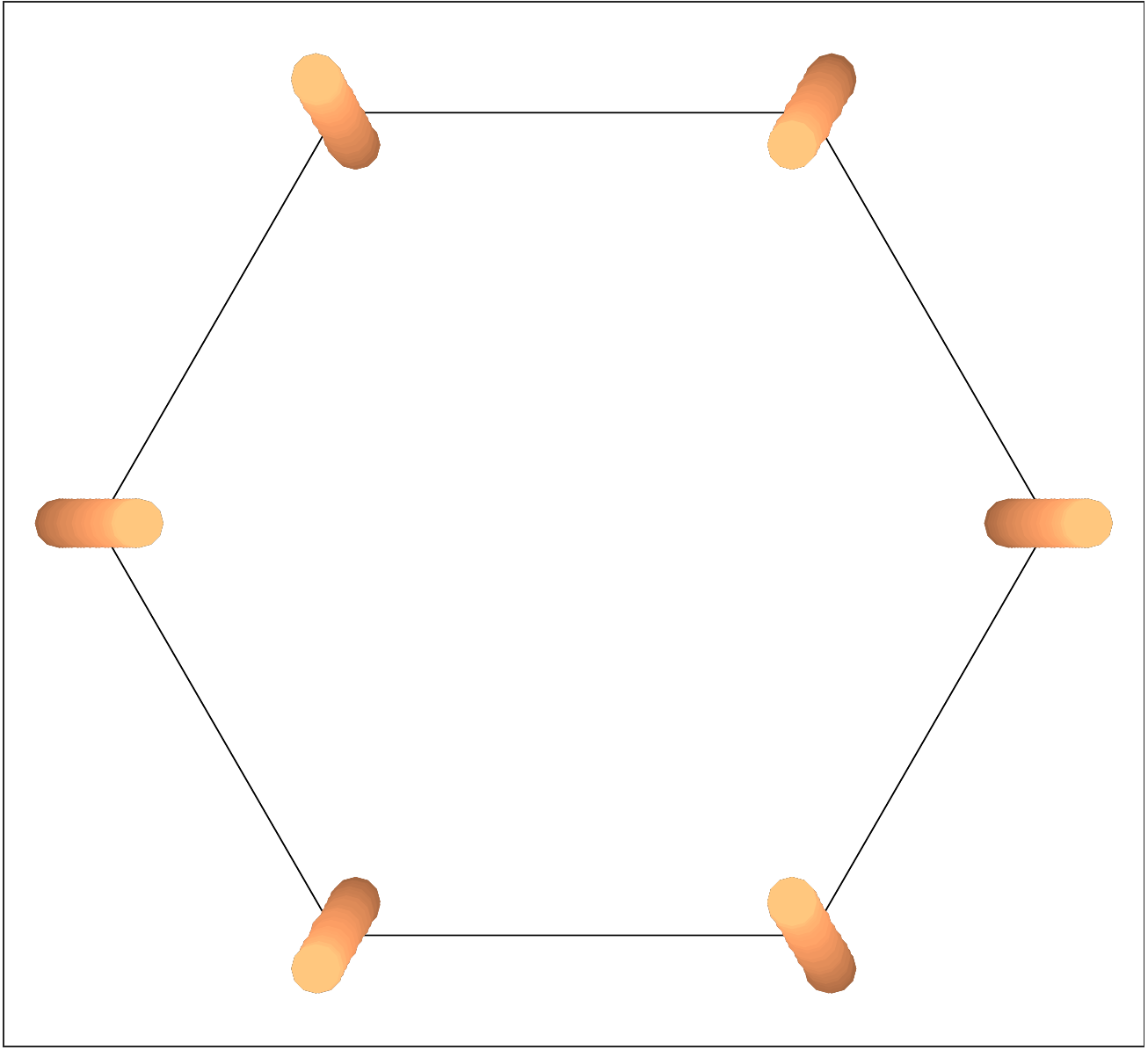} \hspace{3cm} $\Omega_5^\text{(Exp)}$\includegraphics[width=0.2\textwidth]{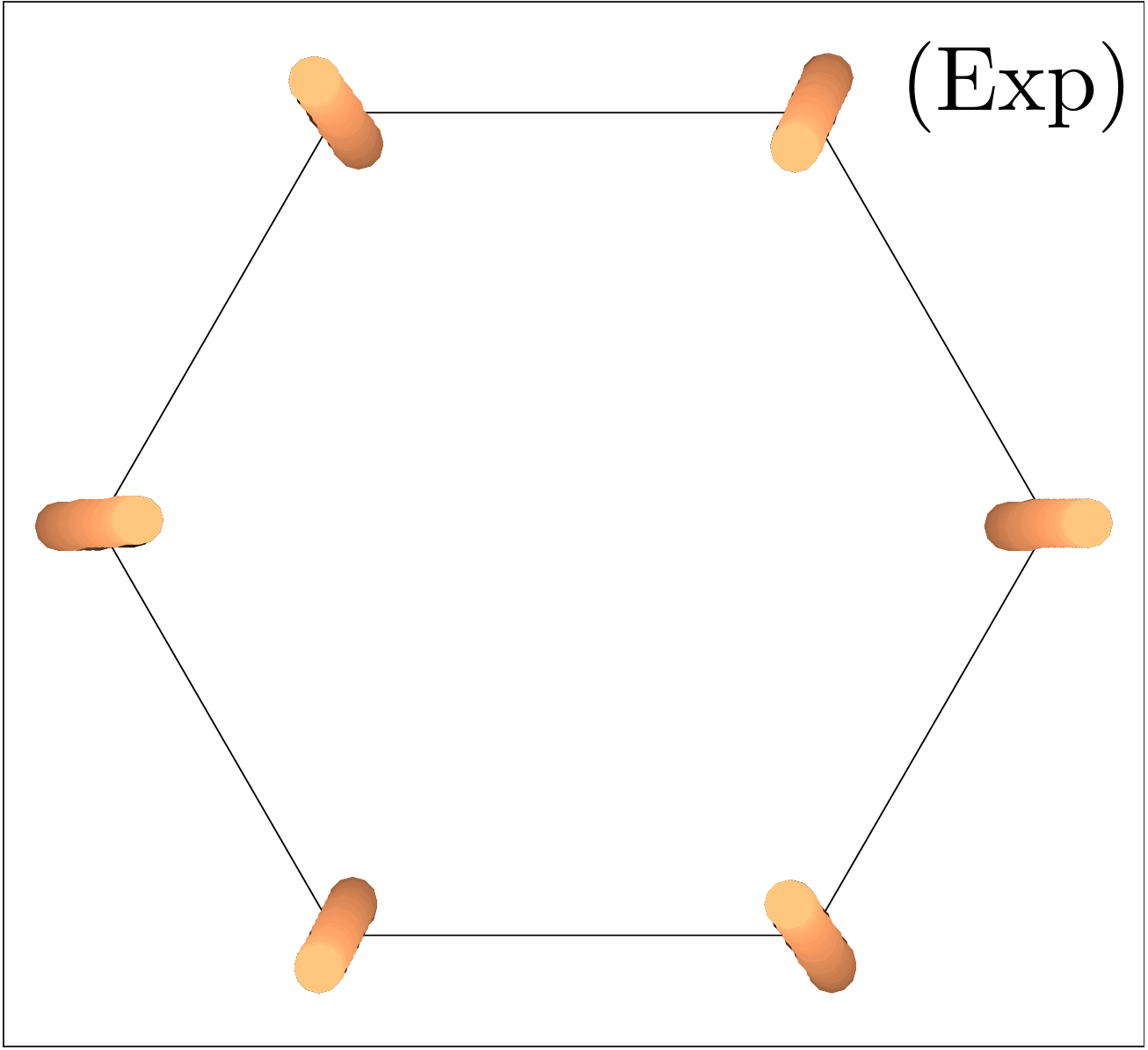}\\
$\Omega_8$ \includegraphics[width=0.2\textwidth]{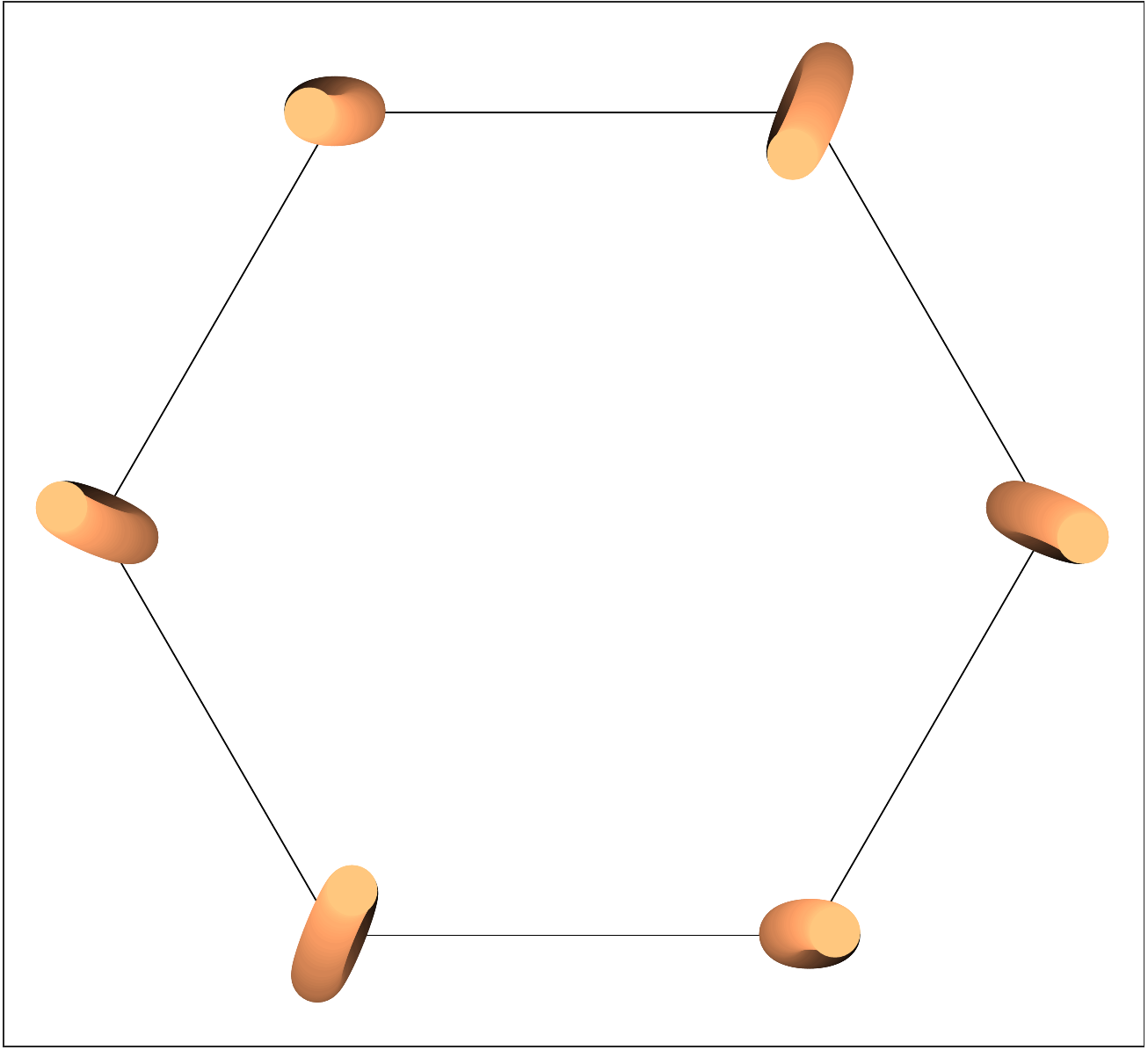}
\includegraphics[width=0.2\textwidth]{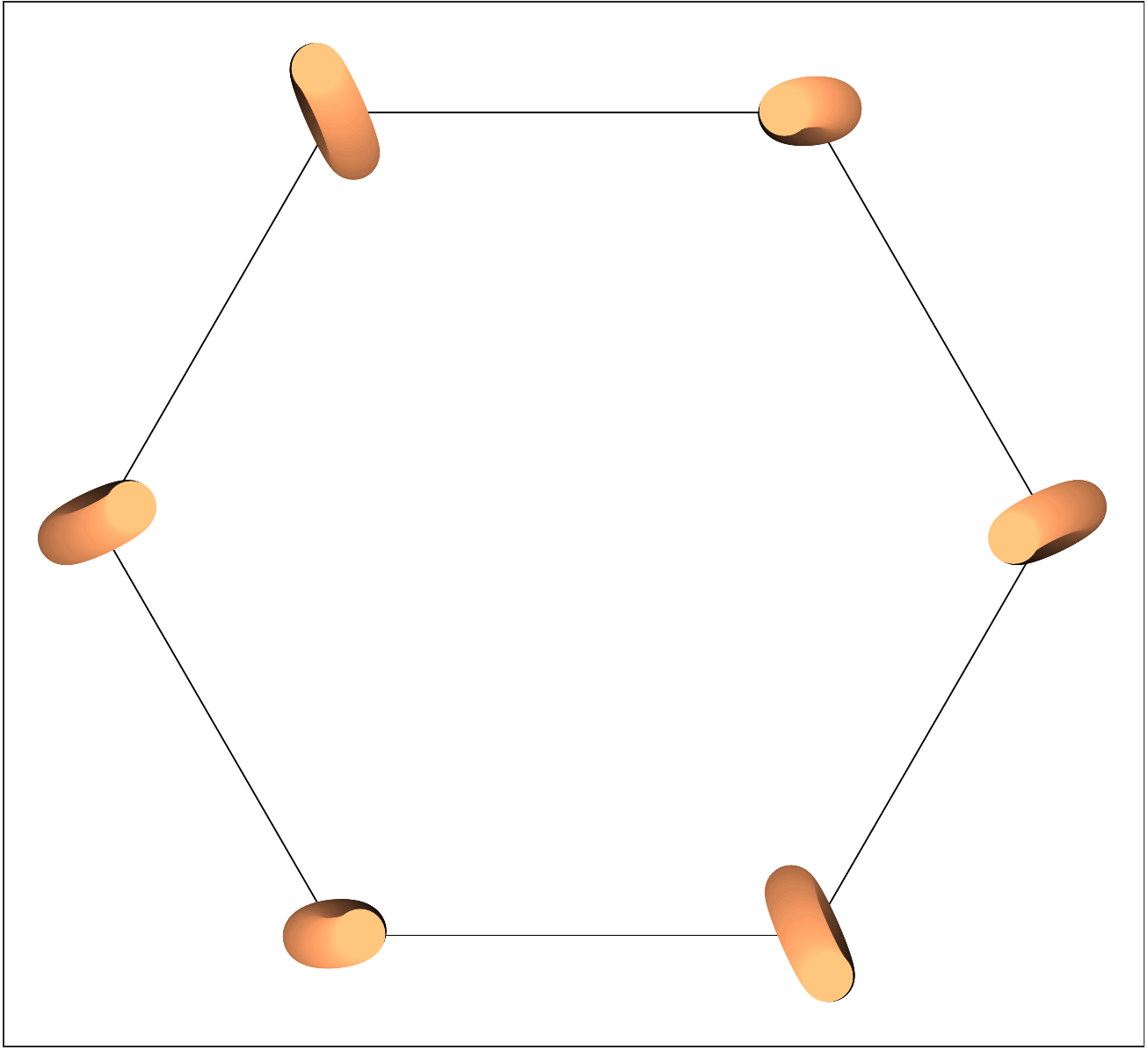} \hspace{1.8cm} $\Omega_8^\text{(Exp)}$\includegraphics[width=0.2\textwidth]{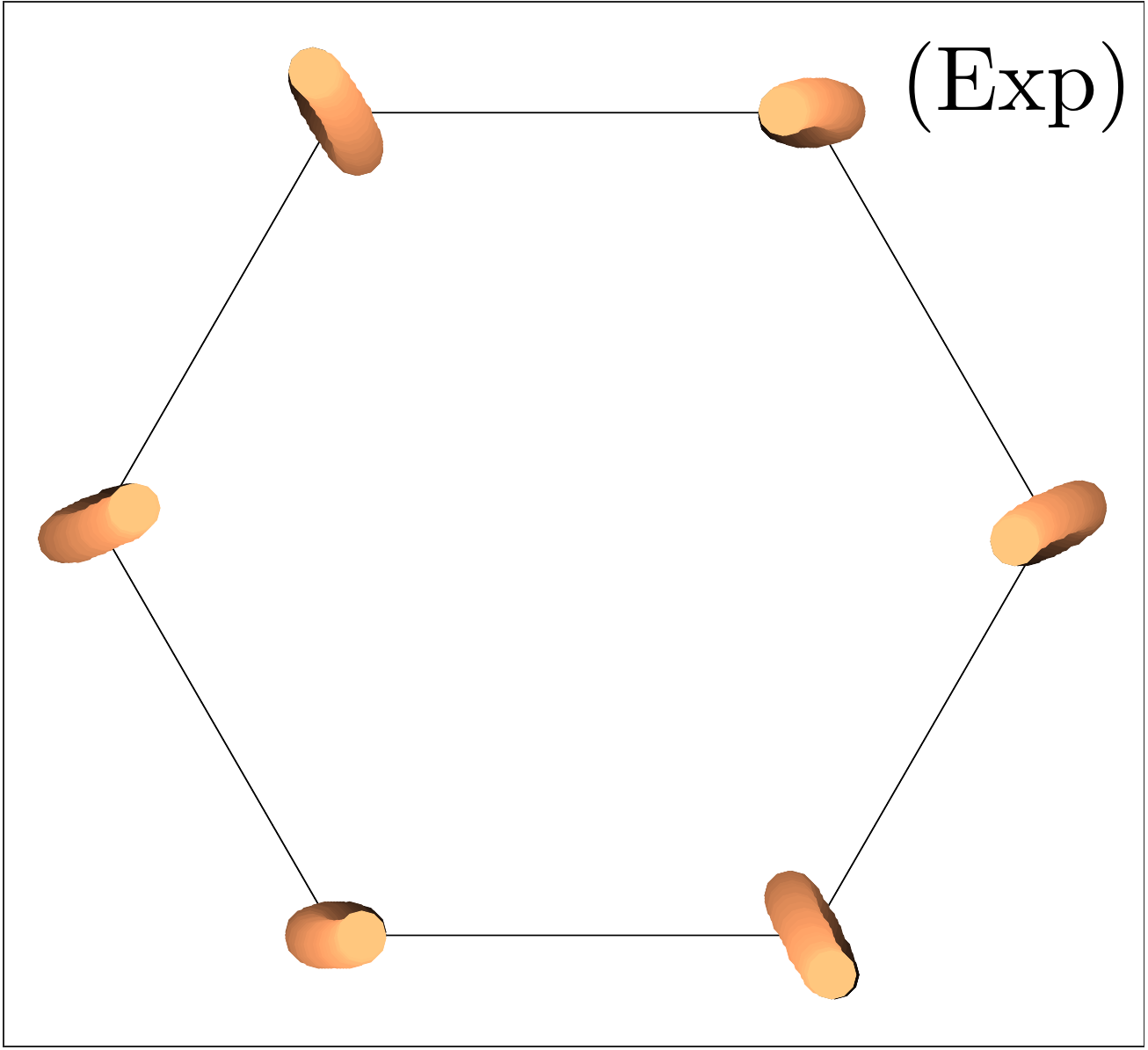}\\
\hspace{1.6cm}$\Omega_3$ \includegraphics[width=0.2\textwidth]{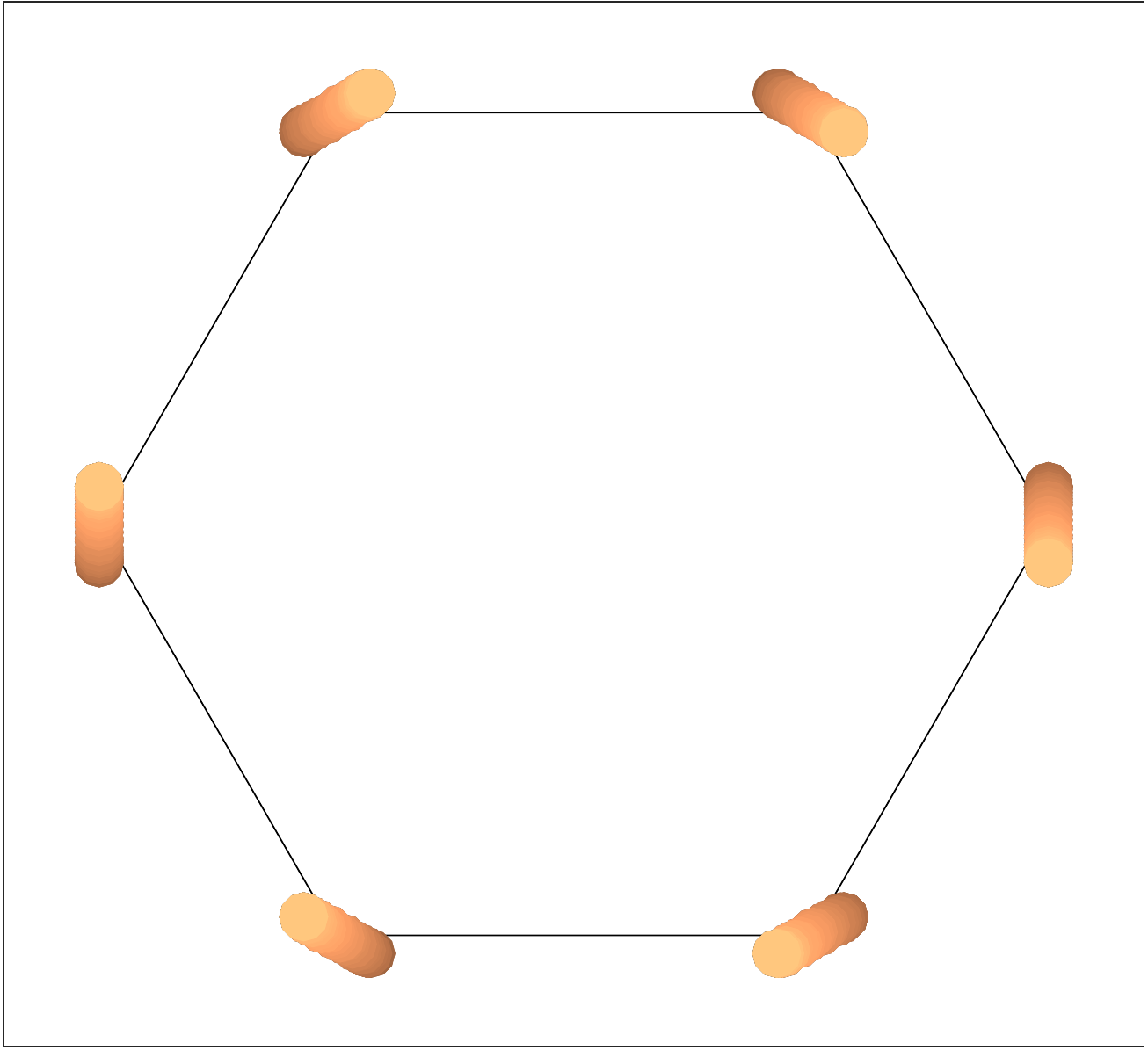} \hspace{3cm}  $\Omega_3^\text{(Exp)}$\includegraphics[width=0.2\textwidth]{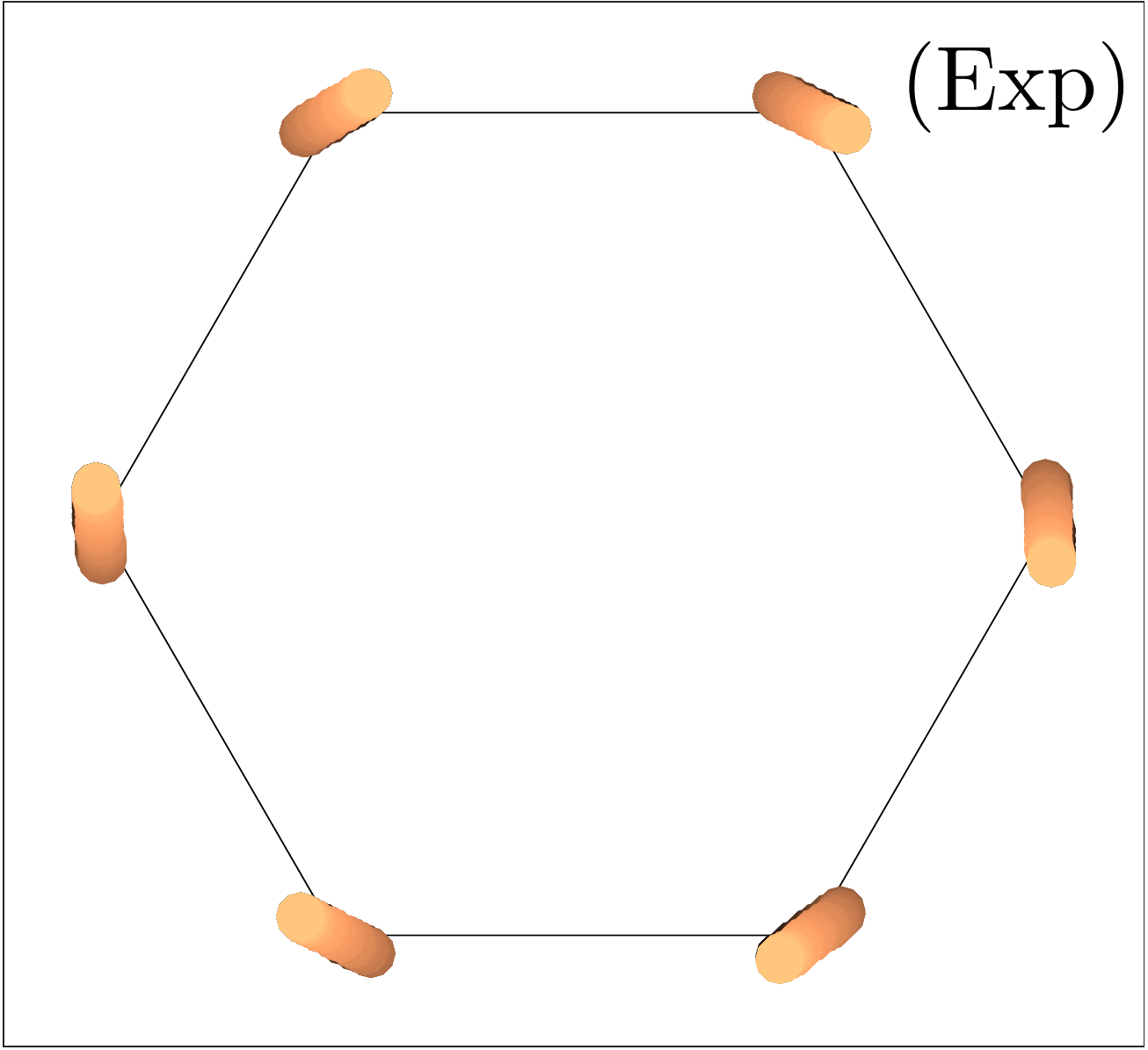}\\
$\Omega_1$ \includegraphics[width=0.2\textwidth]{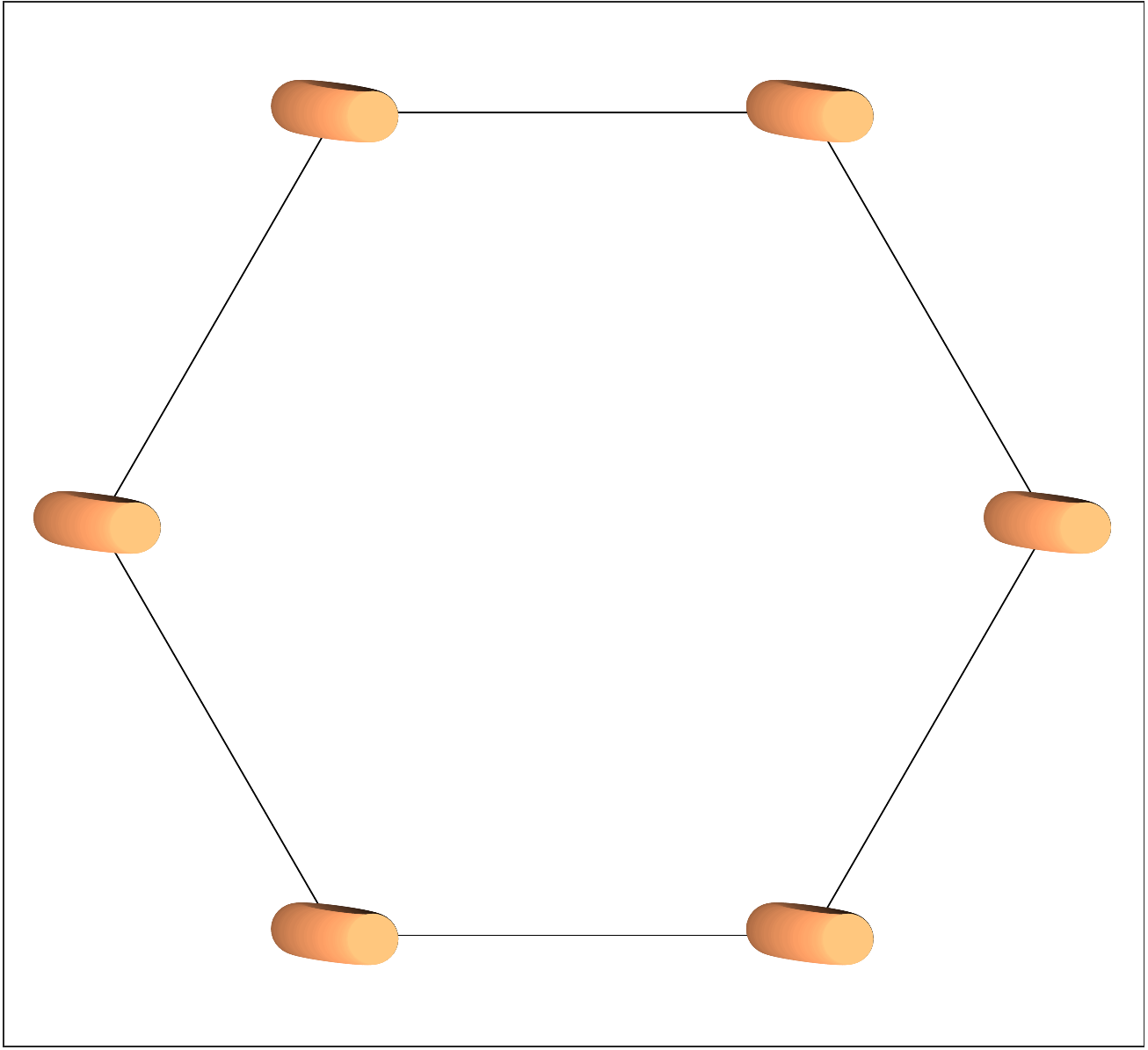}
\includegraphics[width=0.2\textwidth]{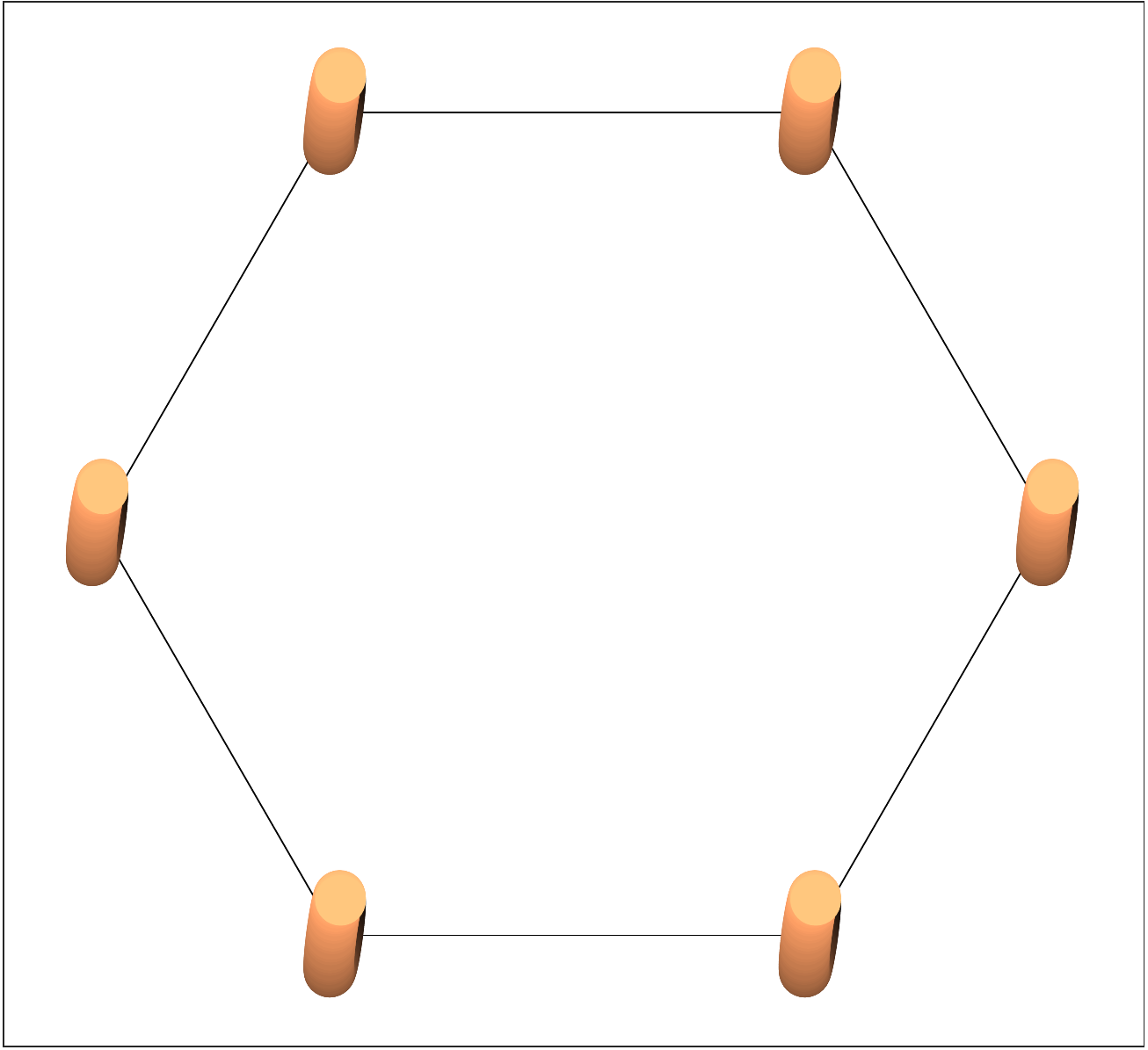} \hspace{1.8cm} $\Omega_1^\text{(Exp)}$\includegraphics[width=0.2\textwidth]{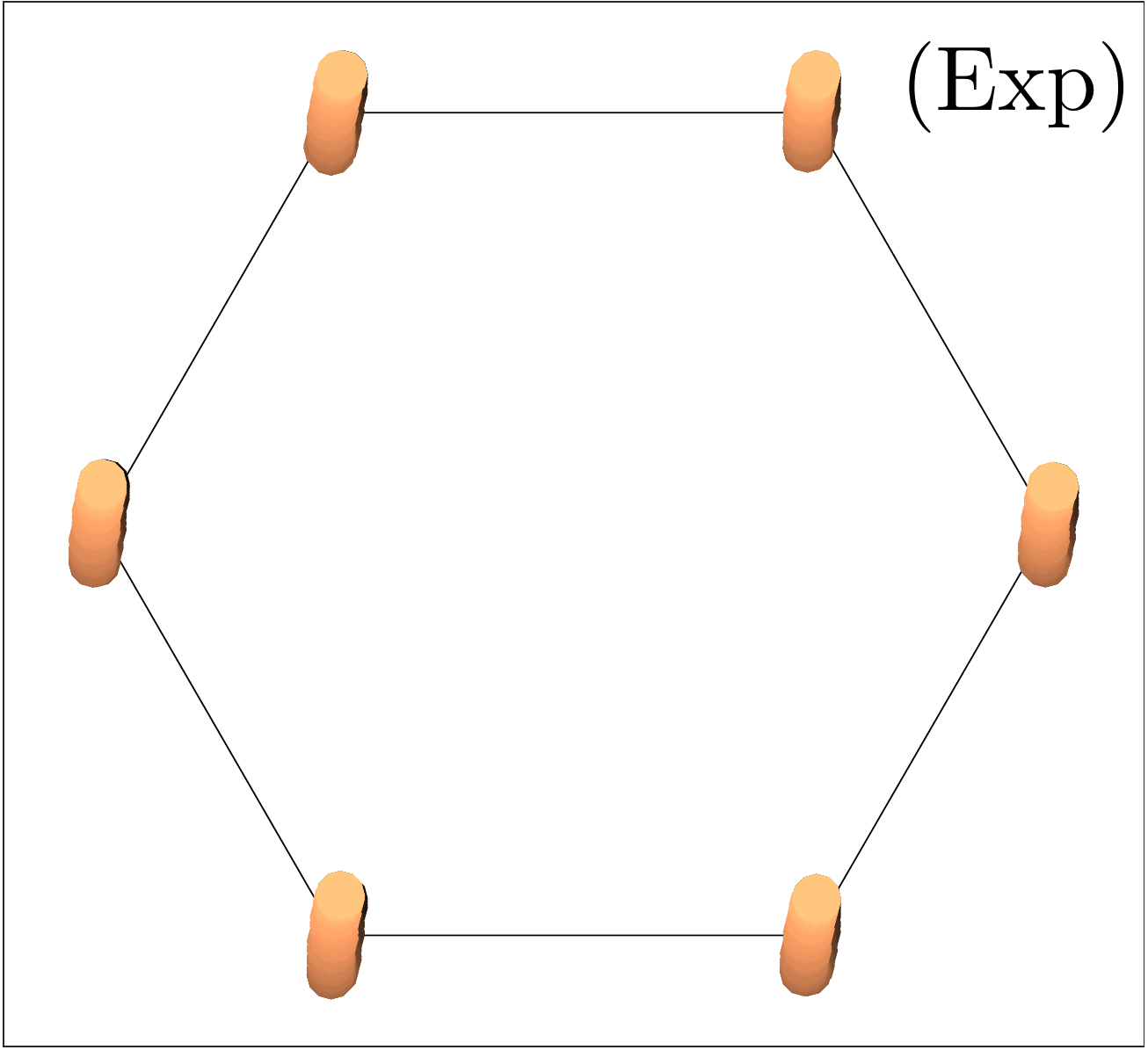}
\caption{Theoretical and experimental spatial oscillation patterns of the eigenmodes of the system for the first configuration of \tref{table:configurations} with $\omega_0=7.2 $~rad$/$s, $\Omega_L= 7.2$~rad$/$s and $\Omega_T= 3.7$~rad$/$s, so that $\Omega_T/\Omega_L=0.52$. From bottom to top, the eigenmodes are ordered according to their increasing frequency of oscillation. The motion of the six pendula around their equilibrium positions is spatially represented within a period of oscillation: The colour gradient is indicative of time, where a darker colour stands for earlier time.
The theoretical modes are the eigenvectors of the matrix $D$ in \eref{dynamicalmatrix}, while the experimental modes are reconstructed from the Fourier transform of the displacements, as discussed in section \ref{sec:results} of the text.}
\label{fig:theoretical_eigenmodes}
\end{figure}

As this total angular momentum $k$ is conserved in our rotationally symmetric system, we can make use of the following Fourier-like transformation
\begin{equation}
\psi_j^S = \frac{1}{\sqrt{6}}\sum_k \e^{-\ii \frac{\pi}{3} \left(k - S\right) j} \varphi_k^S:
\label{angularstates}
\end{equation}
to put the system in \eref{dynamicalmatrix} in a block diagonal form, with each block being a $2$-by-$2$ matrix acting on the sub-space with a given value of $k$. 
The result of this diagonalization exactly recovers the formulas in \eref{eigenmodesLTcoupled} and is graphically illustrated in \fref{fig:angular} and \fref{fig:theoretical_eigenmodes}. Further details can be found in~\cite{Sala_soc}.

Panels (a)-(c) in \fref{fig:angular} show the frequencies of the system as a function of the total angular momentum $k$ for $\Omega_L/\omega_0=1$ for the three cases $\Omega_T/\Omega_L=0,\,0.5,\,1$.
In open dots we show the eigenvalues obtained for the discrete integer values $k \in(-3,3]$, while the solid lines are just guides to the eye. 
The eigenstates, shown as open dots, are labelled according to the analytic expression of their frequencies given in \eref{eigenmodesLTcoupled}: the degenerate states are distinguished by the different value of the total angular momentum, which prevents their mixing  by symmetry as long as the system is rotationally invariant.

The weak value of the spin-orbit coupling in the polariton experiments in~\cite{Sala_soc} restricted that investigation to the case where $|\Omega_T-\Omega_L|\ll \Omega_{L,T}$. This experiment is realized in a micropillar chain making use of two polarization states of the same s-wave orbital state.
In this limit, only four different eigenfrequencies can be clearly spectrally distinguished, as shown in \fref{fig:angular}(c). 
In the mechanical system, this regime is instead hard to access as it requires springs with a very small rest length, $\ell_0/\ell \rightarrow 0$. 

The opposite $\Omega_T/\Omega_L\approx0$ regime is instead realized when the spring rest length is equal to the distance between the pendula $\ell_0=\ell=D$, and there is no pre-tensioning in the equilibrium configuration. 
This regime is illustrated in \fref{fig:angular}(a). 
In polariton systems, this regime is achieved in a lattice of pillars when the L,T states correspond to different orbital p-wave states of micropillars \cite{Amo}.
In this p-wave system, the tunneling amplitude of orbitals aligned orthogonal to the link is in fact strongly suppressed with respect to the one of transverse orbitals. 
The large difference between the coupling amplitudes is apparent in the flatness of the higher p-wave bands of the honeycomb lattice studied in~\cite{Amo}. 

As a general remark, we point out another difference between the polariton system and the classical system of pendula. 
In the polariton system of~\cite{Sala_soc}, the Hamiltonian is chirally symmetric and, as a result, the eigenfrequencies are symmetrically located around the bare cavity frequency. 
Chiral symmetry for our pendula system is instead broken and the coupling induced by springs in systems of pendula always increases the frequency of the oscillation modes. 
This is due to additional terms in \eref{Newton_benzene} where the elastic force acting on the $i$-th pendulum depends on the position $\vec{\psi_{i}}$ of the pendulum itself. 
In the presence of spin-orbit coupling in a hexagonal ring, this results in there being different diagonal elements in the dynamical matrix \eref{dynamicalmatrix} for the x-polarized mode and y-polarized mode of a given site, breaking chiral symmetry.

In \fref{fig:theoretical_eigenmodes} we show the motion of the pendula in each of the twelve eigenmodes.
This is obtained from the eigenvectors of the dynamical matrix $\mathcal{D}$, corresponding to the eigenfrequencies of \eref{eigenmodesLTcoupled}, and its form does not depend on the specific value of $\Omega_T/\Omega_L$.
The panels in the left part of the figure show the theoretical eigenmodes, while in the right part we show a comparison with the experimental modes as obtained from the data analysis. 
The motion of the pendula is represented within a period of oscillation around the equilibrium positions, and the colour gradient represents time, where a darker colour stands for earlier time. 
For the degenerate eigenfrequencies, any linear superposition of two eigenstates of \eref{dynamicalmatrix} can be experimentally observed, with weights determined by the specific excitation procedure. In \fref{fig:theoretical_eigenmodes}, we show a theoretical eigenmode which is constructed to be closest to the experimentally observed one, together with the orthogonal eigenmode. In particular, the right panel of the theoretical degenerate eigemodes corresponds to the experimentally closest mode.
More details on the experimental eigenmodes will be given in the following.

\section{Experimental setup}
\label{sec:experiment}

\begin{figure}[t]
\centering
\includegraphics[height=70mm]{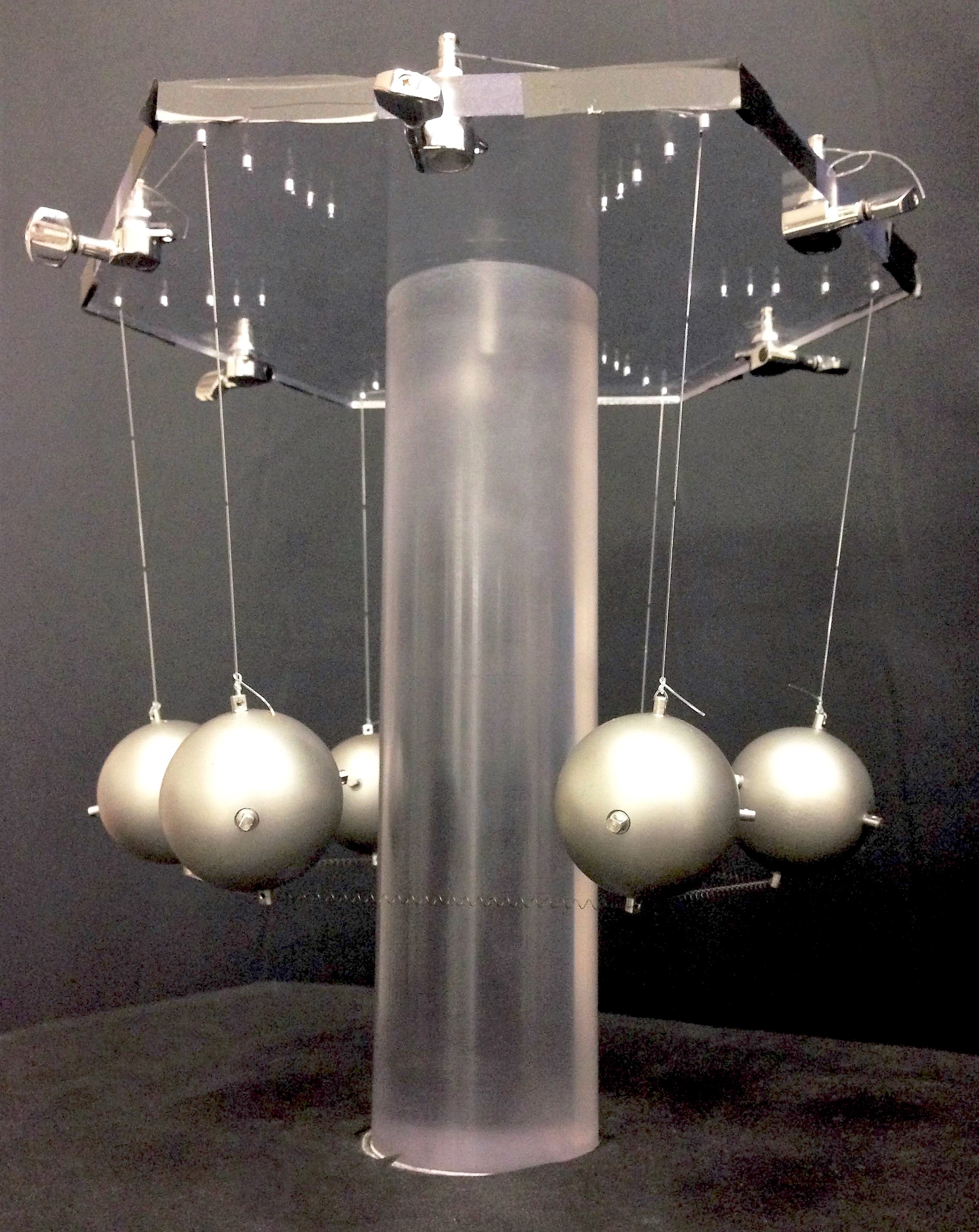}\hspace{1em}
\includegraphics[height=70mm]{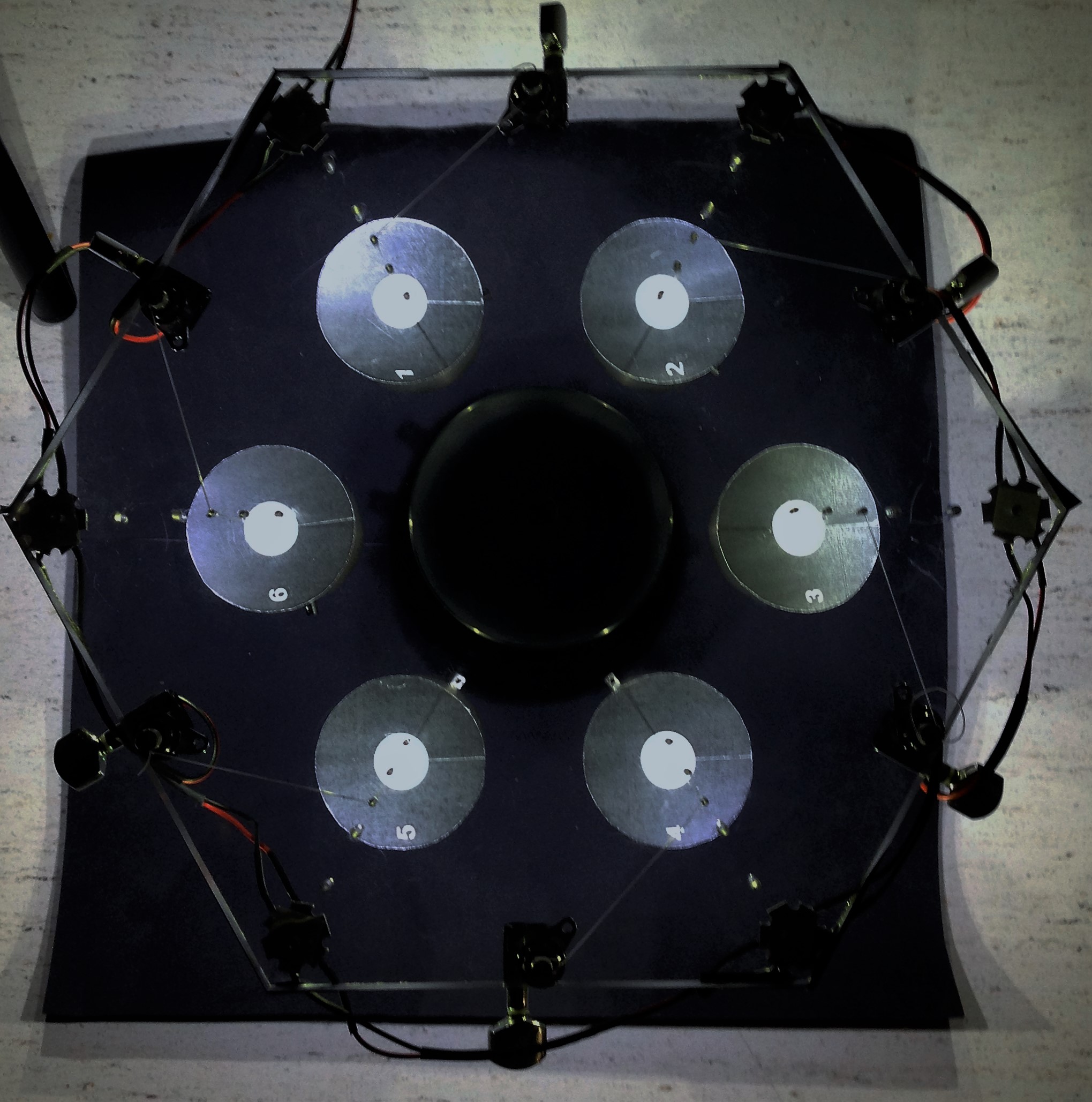}
\caption{On the left, a picture of the setup used for the experiment.
On the right, a snapshot of the video used for measuring the displacements of the pendula through the position of the white circles.}
\label{fig:experimental_setup}
\end{figure}

In~\fref{fig:experimental_setup} 
we show a picture of the experimental setup of six coupled pendula connected with pre-tensioned springs. 
Each pendulum is realized by a sphere of mass  $m=(0.596\pm 0.001)$~Kg and radius $R=(2.65\pm 0.05)$~cm attached to a string of length $L=(16.0\pm0.5)$~cm, hanging from a vertex of the hexagonal transparent plastic roof. 
As visible in \fref{fig:experimental_setup}, the top plastic roof is pierced with sets of holes that allow us to hang the hexagonal system of pendula at five possible distances $D$, as reported in \tref{table:configurations} from the outermost to the innermost.

Under the assumption that the pendula are made of point-like masses attached to a wire of length $L+R$, their natural frequency is expected to be: 
\begin{equation}
\omega_0= \sqrt{\frac{g}{L+R}}=(7.2 \pm 0.1)\text{rad}/\text{s}.
\label{omega0_experimental}
\end{equation}
where we have used the geometrical parameters given above.
The pendula are coupled through springs of rest length $\ell_0=(7.30\pm0.01)$~cm.
The springs are attached to a hook that is located at the bottom of the sphere. This joint between the spring and the bottom hook of the pendulum is flexible, in a sense that it is not glued or soldered, leaving the spring free to rotate.
The distance $D$ between the hanging point of the pendula in the five configurations is always larger than the rest length of the spring and, as a result, the springs are pre-tensioned.
In the equilibrium configuration, the pendula move in with respect to their hanging points, as visible in the left part of \fref{fig:experimental_setup}. 
The values of $\ell$ for the five configurations is reported in \tref{table:configurations}. 

We characterize the springs by measuring their elongations when subjected to a known force and extracting the constant of the spring as the slope of the resulting curve.
We observe that, for extensions $\Delta \ell=(\ell-\ell_0)>15$ mm, the springs behave nonlinearly. However, provided that the extension does not exceed  $\Delta \ell \approx 35$ mm, the elastic response of the springs can still be locally approximated as linear, with a modified spring constant $\kappa^\text{eff}$. When $\Delta \ell>40$ mm, the spring is instead in the plastic regime where the deformations are permanent.
For each of the experimental configurations, the modified value of the spring constant $\kappa^\text{eff}$, which takes into account the nonlinear behaviour, is given in \tref{table:configurations}.
The value of the modified spring constant $\kappa^\text{eff}$ is obtained from a local linear fit of the force-extension curve of the springs in the relevant range of extensions.
In \tref{table:configurations} we also report the corresponding ratio $\Omega_T/\Omega_L$ of the longitudinal and transverse frequencies defined from \eref{ltfrequencies} for the experimental parameters.

\begin{table}[t]
\caption{Parameters for the different configurations used in the experiment. 
  As the distance $D$ between the hanging points of the pendula is reduced, the length $\ell$ of the spring in the equilibrium configuration decreases. 
  The value of the spring stiffness $\kappa^\text{eff}$ includes nonlinear effects and is obtained from a linear fit of the force-extension curve of the springs.}
\begin{indented}
\item[]\begin{tabular}{c c c c c}
Configuration	& $D$ [mm] & $\ell$ [mm] & $\kappa^\text{eff}$ [N/m] & $\Omega_T/\Omega_L$\\
\hline
i						& $142.0 \pm 0.2$ 	& $100.0\pm 0.2$		& $31 \pm 1$	 	& $0.52 \pm 0.03$\\
ii 						& $124.0\pm 0.2$  	& $92.5\pm 0.2$		& $30 \pm 1$	 	& $0.46 \pm 0.03$\\
iii           			& $105.0 \pm 0.2$ 	& $85.0\pm 0.2$  		& $28\pm 1$	 	& $0.37 \pm 0.04$\\
iv           			& $97.5 \pm 0.2$		& $81.0\pm 0.2$		& $27\pm 1$	 	& $0.31 \pm 0.05$\\
v     					& $87.0\pm 0.2$    	& $77	.0\pm 0.2$  	& $27\pm 1$	 	& $0.23 \pm 0.08$\\
\hline
\end{tabular}
  \label{table:configurations}
\end{indented}
\end{table}

To excite the system, we displace by hand one of the pendula from its equilibrium position and suddenly release it, as shown in the supplementary video.
The local nature of the initial condition means that the subsequent motion of the pendula is a superposition of all the eigenmodes of the system depicted in \fref{fig:theoretical_eigenmodes}. 
A different choice in the direction of the initial condition would only affect the relative weight of each mode.
At late times, we notice that the oscillations damp out as one would expect under the effect of friction. The characteristic time-scale is of the order of $95$~s, which means about $110$ periods of oscillations. In particular, it seems that the dominant contribution to friction does not come from the springs, but rather from the friction at the pivot point between the roof and the string, and also from air resistance.

\begin{figure}[t]
\centering 
\includegraphics[width=0.95\textwidth]{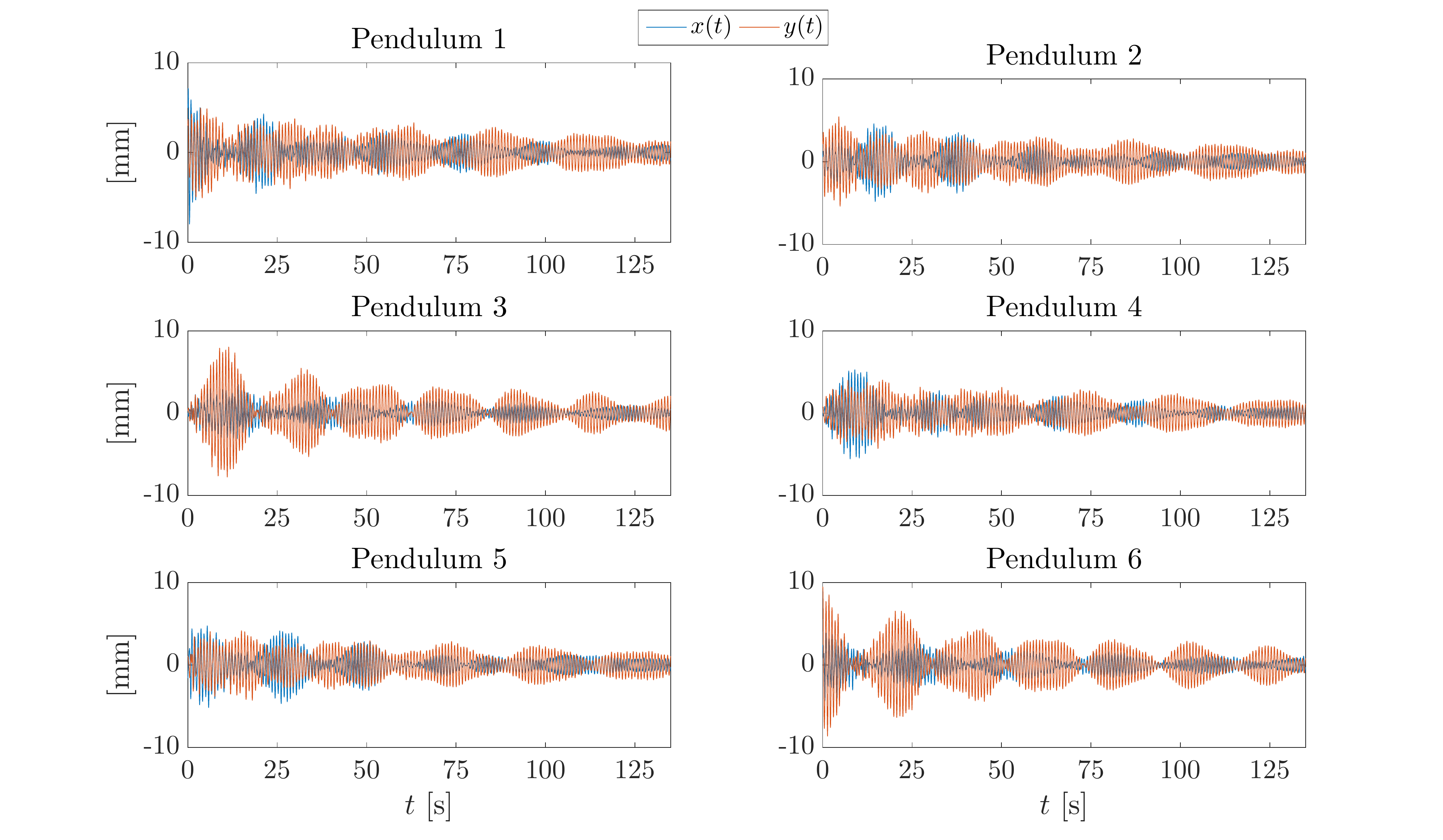}
\caption{Experimental displacements $x_i(t)$ and $y_i(t)$ of the six pendula from their equilibrium positions as extracted from the video analysis for the fifth configuration in \tref{table:configurations},  $\omega_0= 7.2$rad$/$s, $\Omega_L= 7.2$rad$/$s and $\Omega_T= 1.6$rad$/$s.}
\label{fig:experimental_motion}
\end{figure}

The motion of the pendula is recorded with a standard video camera, positioned above the system.
This motion is sampled at a frequency $\nu_s=25$~Hz, for a total time $T$ of about $120$~s.
The video is then digitally analysed to obtain the displacements of the centre of mass of each pendulum.
In order to facilitate the measurement of the position of the pendula, white circles of paper have been rigidly attached to the top of the spheres, as shown in the right panel of \fref{fig:experimental_setup}, see the supplementary video.
In \fref{fig:experimental_motion} we show a typical experimental result for the motion of the six pendula.
We then perform a temporal Fourier transform on each set of data to obtain the Fourier amplitudes for both the $x$ and the $y$ components of each pendulum: $|F_{x_i}(\Omega)|$ and $|F_{y_i}(\Omega)|$. 
The frequencies span from $\Omega \in [-\pi \nu_s, \pi \nu_s ]$, with a step of $\Delta\Omega=2\pi/T$.

\section{Results and discussions}
\label{sec:results}

\begin{figure}[t]
\centering
\includegraphics[width=0.45 \textwidth]{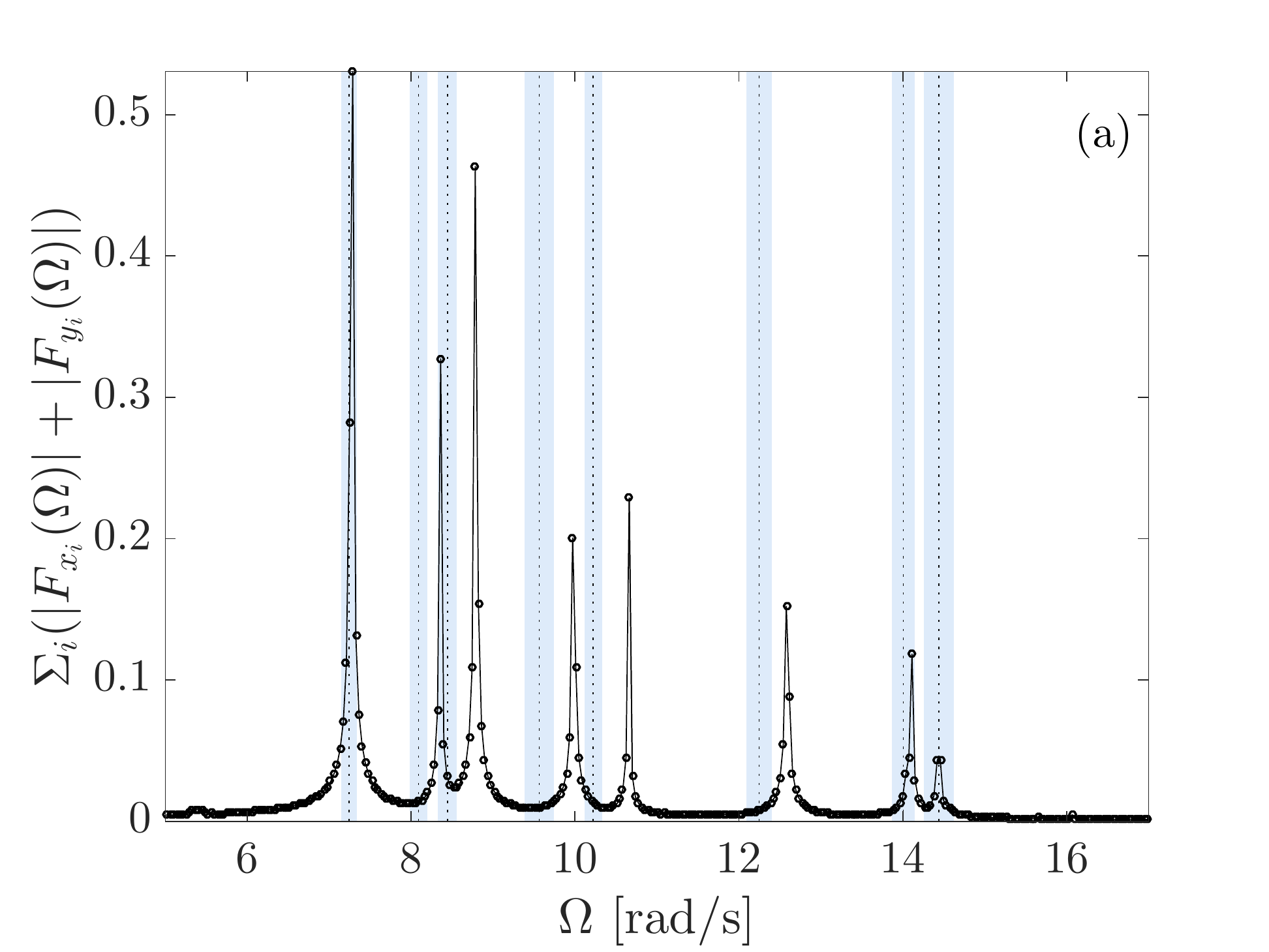}
\includegraphics[width=0.45 \textwidth]{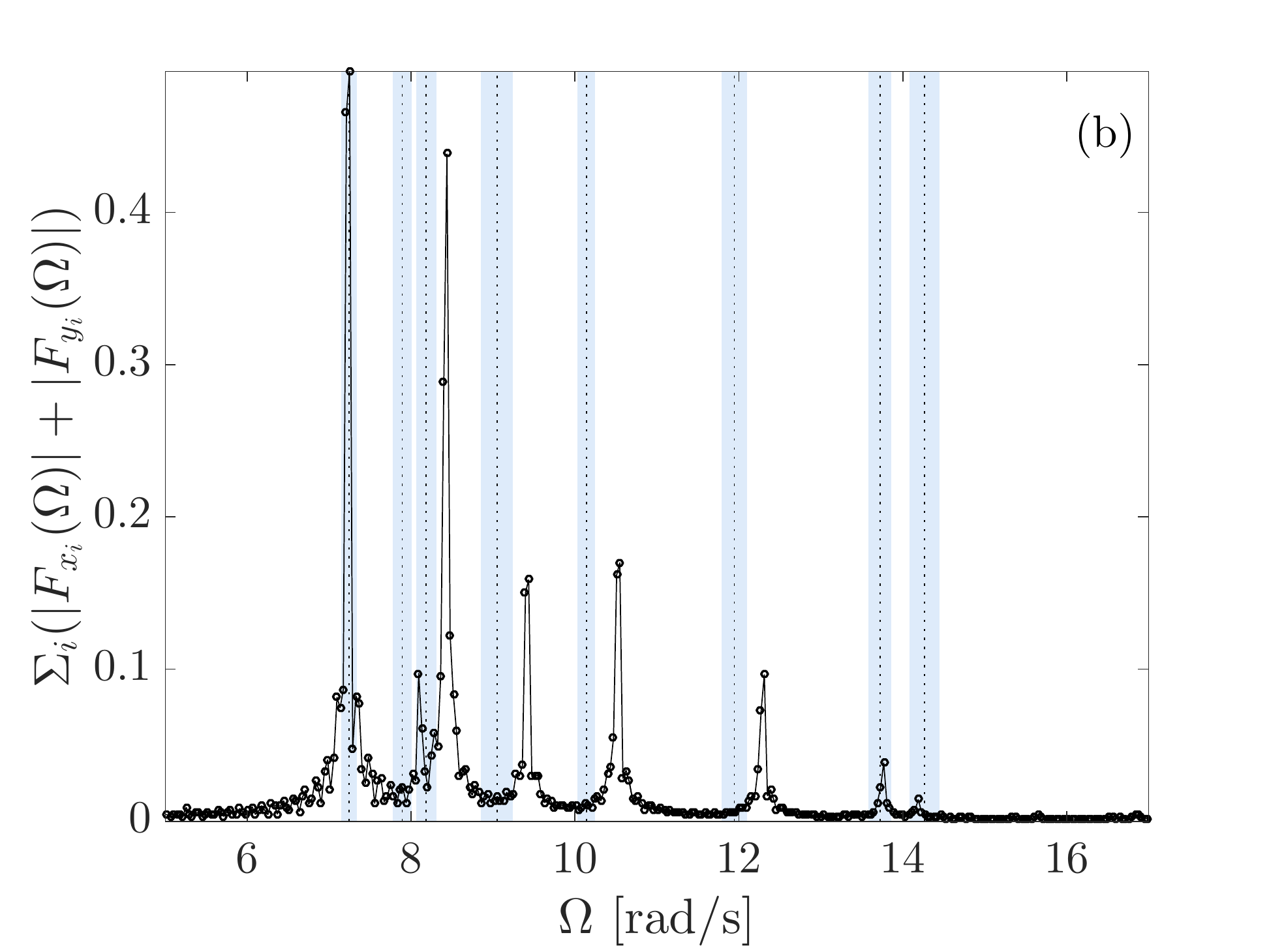}\\
\includegraphics[width=0.45 \textwidth]{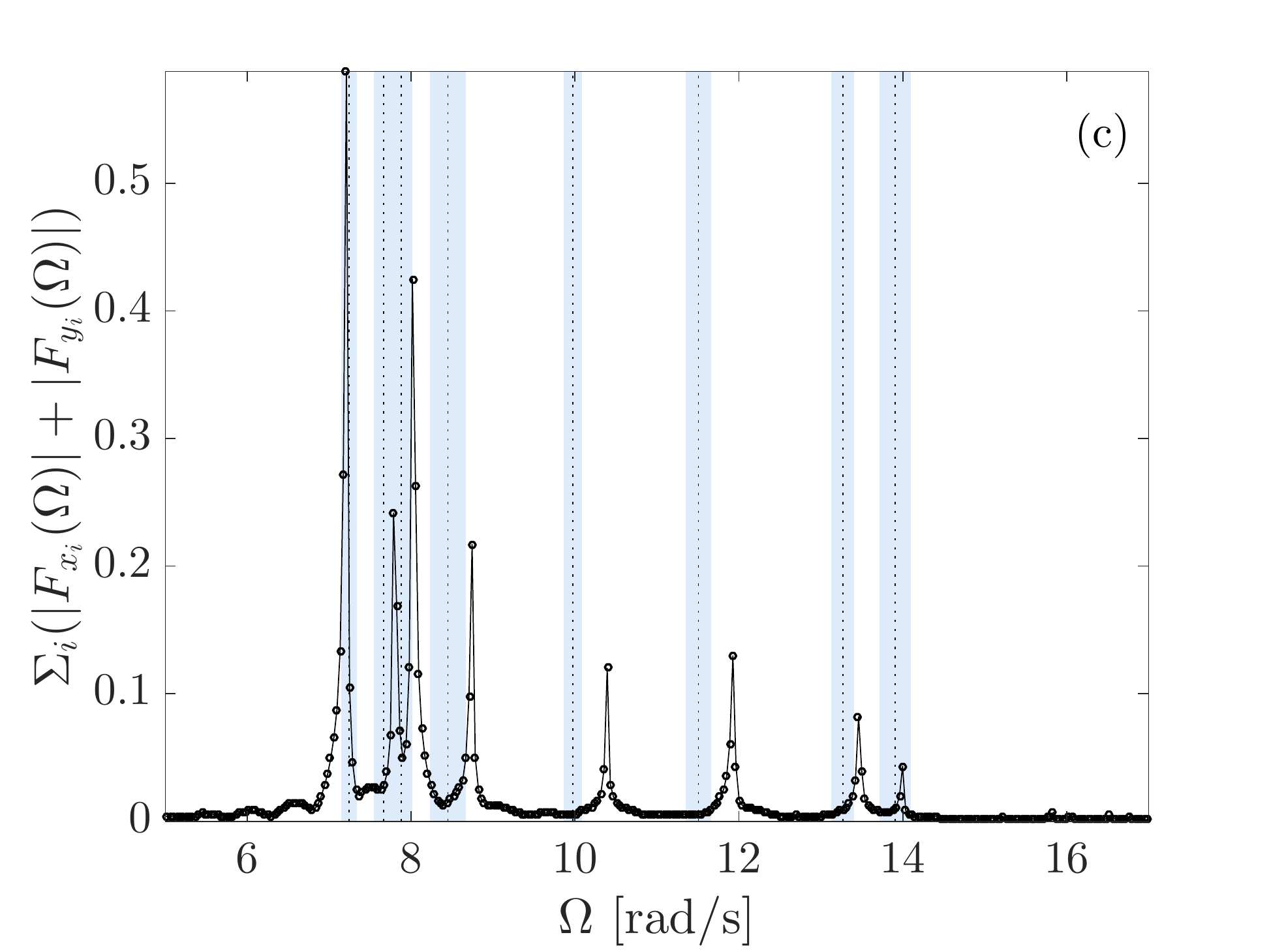}
\includegraphics[width=0.45 \textwidth]{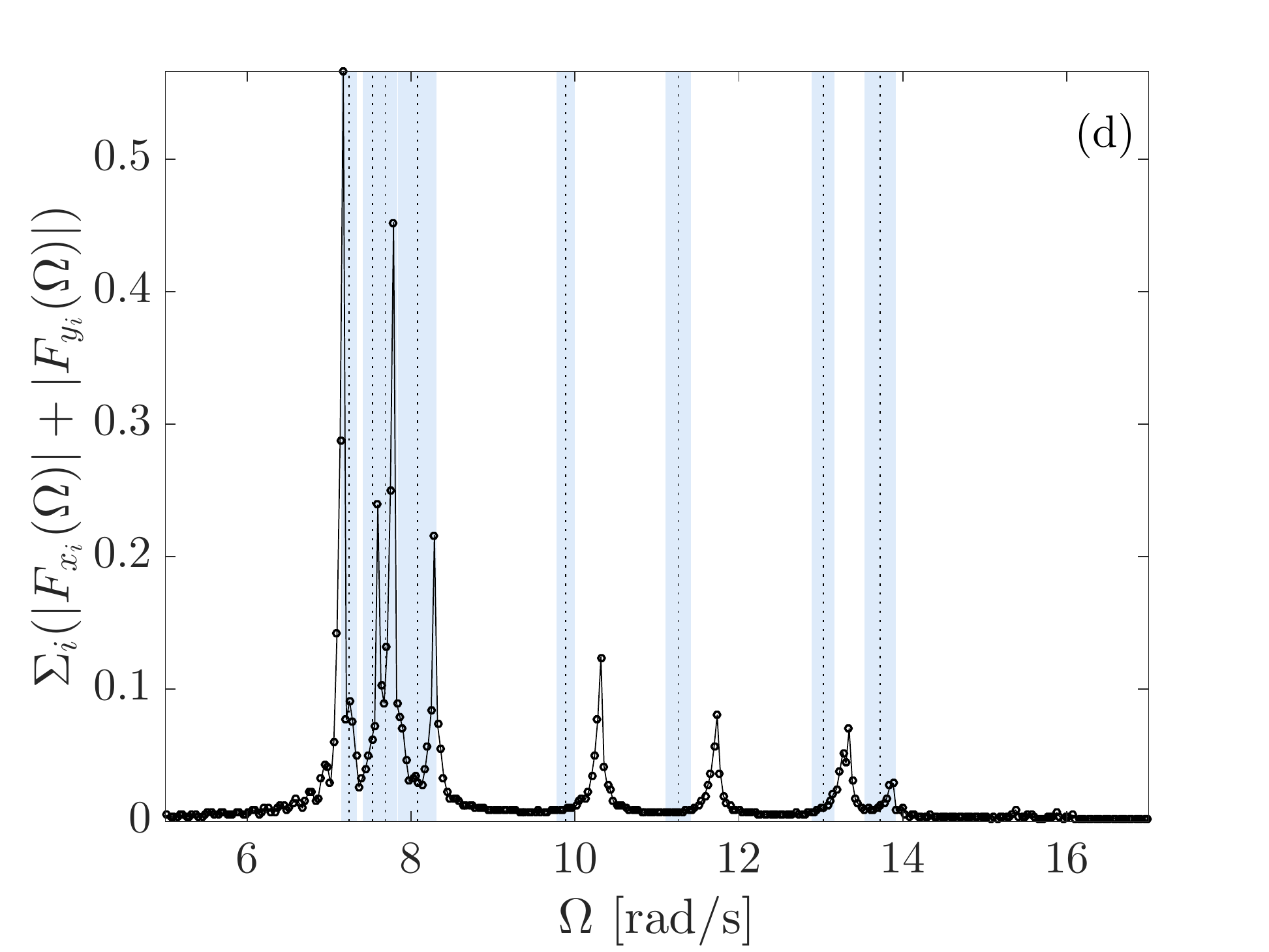}\\
\includegraphics[width=0.45 \textwidth]{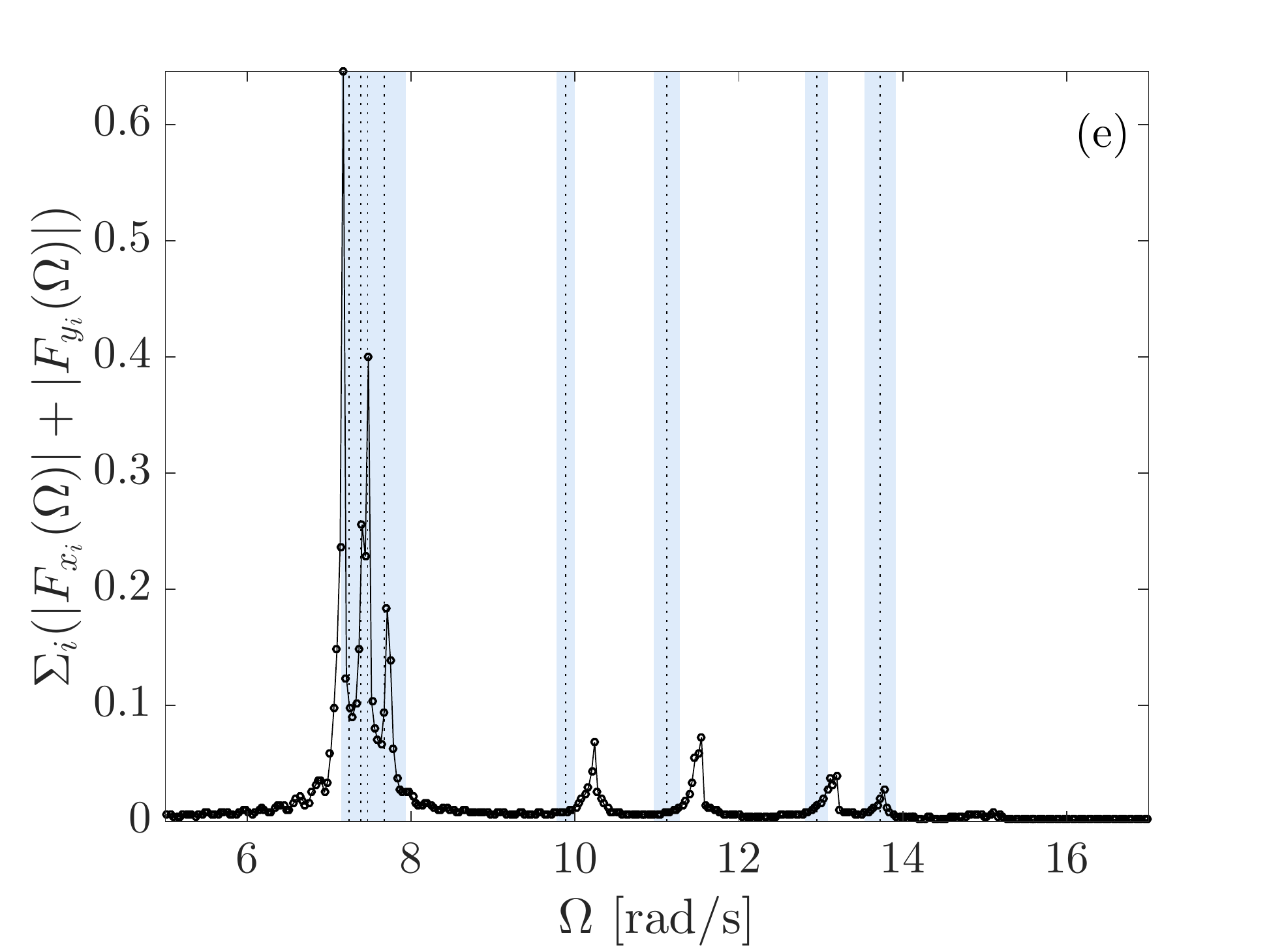}
\caption{Experimental spectra for the five different configurations of \tref{table:configurations}, corresponding to different pre-tensioning.
Panels from (a) to (e): $\Omega_T/\Omega_L= 0.52,\,0.46,\,0.37,\,0.31,\,0.23$ and $\omega_0= 7.2$rad$/$s. 
Dots show the normalised total amplitude of the Fourier spectra, $\sum_{i=1}^6 \left(|F_{x_i}(\Omega)|+|F_{y_i}(\Omega)|\right)$, while the solid line is a guide to the eye. 
The black dashed lines correspond to the theoretical predictions~\eref{eigenmodesLTcoupled} for the eigenfrequencies within the simple pendula approximation. The light blue areas indicate the experimental errors.}
\label{fig:exp_spectra_conf}
\end{figure}

In \fref{fig:exp_spectra_conf} we show the frequency spectra as calculated from the total Fourier amplitude, for frequencies in the region of interest. 
Each spectrum has been normalised in such a way that the integral over the whole frequency range is equal to one.
Panels (a)-(e) of \fref{fig:exp_spectra_conf} show the experimental results for different values of the pre-tension of the springs as summarized in \tref{table:configurations}.
Black dots are the experimental spectra, obtained as $\sum_{i=1}^6 \left(|F_{x_i}(\Omega)|+|F_{y_i}(\Omega)|\right)$.
The theoretical eigenfrequencies, as calculated from \eref{eigenmodesLTcoupled} within the simple pendula approximation using the experimental values of \tref{table:configurations}, are shown with dashed black vertical lines in panels (a)-(e) of \fref{fig:exp_spectra_conf}. 
The light blue areas around the theoretical eigenfrequencies indicate the errors associated to the frequencies in \eref{eigenmodesLTcoupled} calculated from the experimental uncertainties in the parameters. 
The relative height of the peaks depends on the initial condition. 
A different initial condition will excite another superposition of modes, each of them with a different coefficient and hence with different spectral heights, but the frequencies of the peaks are not affected by the initial condition.

In \fref{fig:exp_spectra_conf}, we observe a qualitative overall agreement between the experimental spectra and the theoretical predictions. 
In particular the low frequency peaks in \fref{fig:exp_spectra_conf}(c)-(e) match the theoretical predictions well within the experimental error. 
On the other hand, appreciable deviations are visible for the peaks around $11$ rad/s for all the panels. 
A convincing explanation for these discrepancies will be given in the next subsection.

Before entering into this discussion, it is useful to look at the Fourier transform of the displacements, which allows us to reconstruct the oscillation amplitude pattern of the eigenmodes from the $F_i(\Omega)$ evaluated for $\Omega$ located at a peak. 
The experimental eigenmodes are plotted in the right part of \fref{fig:theoretical_eigenmodes}, labelled as (Exp) and ordered, from bottom to top, according to the increasing value of the corresponding eigenfrequency. 
The oscillation patterns of the experimental eigenmodes are in excellent agreement with the ones of the theory depicted in the left part of \fref{fig:theoretical_eigenmodes} and discussed in section \ref{sec:hexagon}, especially regarding the symmetry of the pattern.

In particular, we notice that the eigenmodes associated with eigenfrequencies $\Omega_3$ and $\Omega_4$ present an azimuthal symmetry for the pattern of oscillation, while the eigenmodes associated with eigenfrequencies $\Omega_2$ and $\Omega_5$ present a radial symmetry pattern. 
Such well-defined radial and azimuthal patterns are peculiar of spin-orbit coupled systems~\cite{Sala_soc}.

\begin{figure}[t]
\centering 
\includegraphics[width=0.8\textwidth]{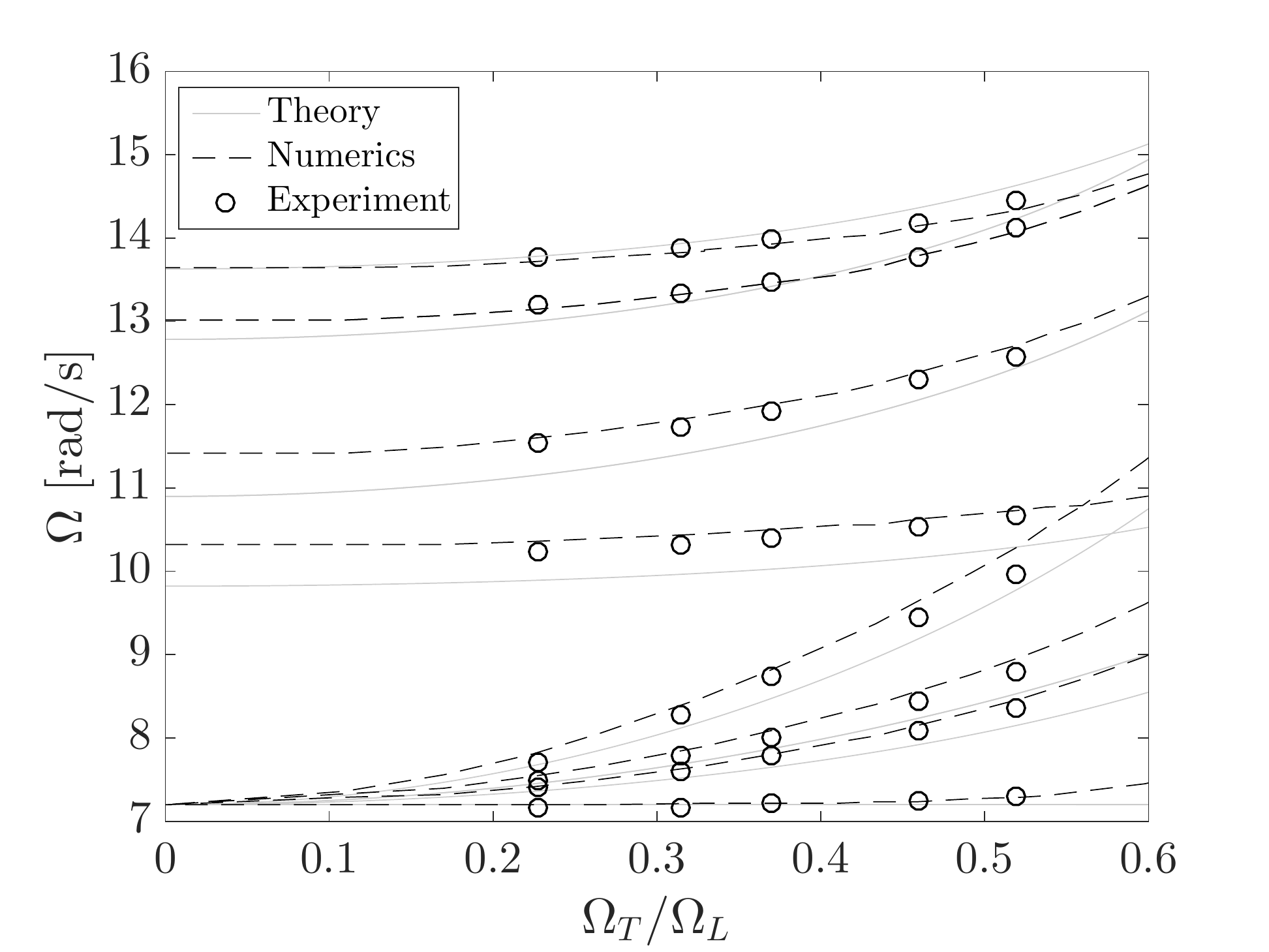}
\caption{Frequency of the eigenmodes of the hexagonal ring of pendula as a function of the ratio $\Omega_T/\Omega_L$.
The open dots are the experimental eigenmodes obtained as the peaks in the frequency spectra for different pre-tensioning of the springs.
Solid lines are the frequencies as obtained from the simple model for point masses in \eqref{eigenmodesLTcoupled}, while dashed lines are obtained from a numerical simulation that includes a non-zero radius of the masses of the pendula.
Both lines are calculated for the experimental parameters, including corrections for the nonlinearity of the springs.
The order of the curves, from bottom to top on the right side, is: $\Omega_1,\,\Omega_3,\,\Omega_8,\,\Omega_2,\,\Omega_5,\,\Omega_6,\,\Omega_7,\,\Omega_4$.
}
\label{fig:SOC}
\end{figure}

\subsection{Comparison with an upgraded model}

As a last point, we wish to shine light on the deviations visible in \fref{fig:exp_spectra_conf} between the theoretical frequencies derived in section \ref{sec:model} and the experimental spectra. 
We can interpret them as a consequence of the finite size of the sphere making up the pendulum and, more precisely, of its rotation around the hook that connects it to the string. 
Such an additional degree of freedom is in fact not included in the simple model discussed in section \ref{sec:model} and gives extra oscillation modes at higher frequency. 
To verify this hypothesis, we numerically simulated the system by solving Euler-Lagrange equations that take into account the effect of a non-zero radius $R$ of the mass and its rotation around the hook. More details on such an approach are given in \ref{appendix2}.

As a first consistency check, we have verified that the full numerical simulation well reproduces the eigenmodes \eref{eigenmodesLTcoupled} in the limit of $R\rightarrow 0$. 
As the rotational symmetry of the system is the same in the extended approach, we expect that the symmetry of the mode oscillation patterns is unchanged: this has also been successfully checked for different values of $R$.

Finally, when $R$ is taken to be equal to the actual experimental value, the resulting eigenfrequencies are found in excellent agreement with the experimental ones. 
This is displayed in detail in~\fref{fig:SOC}, where we plot the frequencies of the eigenmodes as a function of the ratio $\Omega_T/\Omega_L$.
Dots are the experimental eigenmodes as obtained from the peaks in the spectra of \fref{fig:exp_spectra_conf} for different hanging positions and therefore different values of the spring pre-tensioning. 
Solid lines are obtained for the theoretical frequencies in \eqref{eigenmodesLTcoupled} for the simple pendulum model involving point-like masses using experimental parameters. 
Dashed lines are instead the eigenfrequencies obtained from the full numerical simulation including the non-zero radius of the masses and their rotation around the hook. 
An excellent agreement is found over a wide range of $\Omega_T/\Omega_L$, \textit{i.e.} of spin-orbit coupling strength.

\section{Conclusions}
\label{sec:conclusions}

In this paper we have given experimental evidence for a tunable spin-orbit coupling in classical mechanics using a system of six coupled pendula arranged in a hexagonal geometry and connected by pre-tensioned springs. 
The experimental results for the oscillation frequencies and the oscillation patterns are compared with theoretical models: while the qualitative agreement with a simple pendulum approximation is already quite good, it becomes quantitatively excellent once we take into account the finite radius of the masses and their rotation around the hook connecting them to the string. 
By changing the hanging position of the pendula, we have demonstrated how the strength of the spin-orbit coupling is tunable just by varying the amount of pre-tensioning of the springs.

Future developments will include extending the experimental study of the spin-orbit coupling to larger two-dimensional lattices of pendula. 
Such an extension can be achieved by simply connecting more springs and pendula to form a honeycomb lattice, with no additional fundamental difficulties.
In this case one could study the topological Lifshitz transition as theoretically proposed in \cite{Kariyado}, and observe the creation, motion and annihilation of Dirac cones.
Moreover, the simulation of an artificial magnetic field for this simple mechanical system is straightforward to realise experimentally by mounting the system on a rotating table~\cite{Kariyado,WangCoriolis} or by adding a spatially inhomogeneous strain in a honeycomb geometry~\cite{Grazia}, so to study the interesting interplay of orbital magnetic effects with the spin-orbit coupling induced by the pre-tensioned springs. \\

\textit{Note:} Immediately before the submission of the paper, a related theoretical work appeared on spin-orbit coupling in mechanical graphene \cite{WangSOC}.

\ack
We thank Gabriele Ferrari, Alberto Amo, and Jacqueline Bloch for fruitful discussions on different aspects of this work.
We acknowledge Giuseppe Vettori for his work during an early stage of the experiment.
This work was supported by the ERC through the QGBE grant, by the EU--FET Proactive grant AQuS, Project No. 640800, and by the Autonomous Province of Trento, partially through the project SiQuro. 
H.M.P. was also supported by the EC through the H2020 Marie Sk\l{}odowska--Curie Action, Individual Fellowship Grant No: 656093 ``SynOptic''.\\
N.M.P. is supported by the European Research Council (ERC PoC 2015 SILKENE No. 693670) and by the European Commission H2020 under the Graphene Flagship (WP14 “Polymer composites,” No. 696656) and under the FET Proactive (“Neurofibres” No. 732344).

\appendix

\section{More details on the upgraded model}

In this appendix we give more details on the derivation of the Euler--Lagrange equations that were used to numerically integrate the dynamics of the system and obtain the results shown in \fref{fig:SOC}.

We include in our theory the effects of a non-zero radius of the sphere making up the pendula. 
In particular, we have to consider that the centre of mass of the sphere does not coincide neither with the top hook where the string is attached, nor with the bottom hook where the coupling springs are connected. These three points are instead aligned along a direction which defines the axis of the sphere. 

\begin{figure}
\centering
\includegraphics[width=0.4\textwidth]{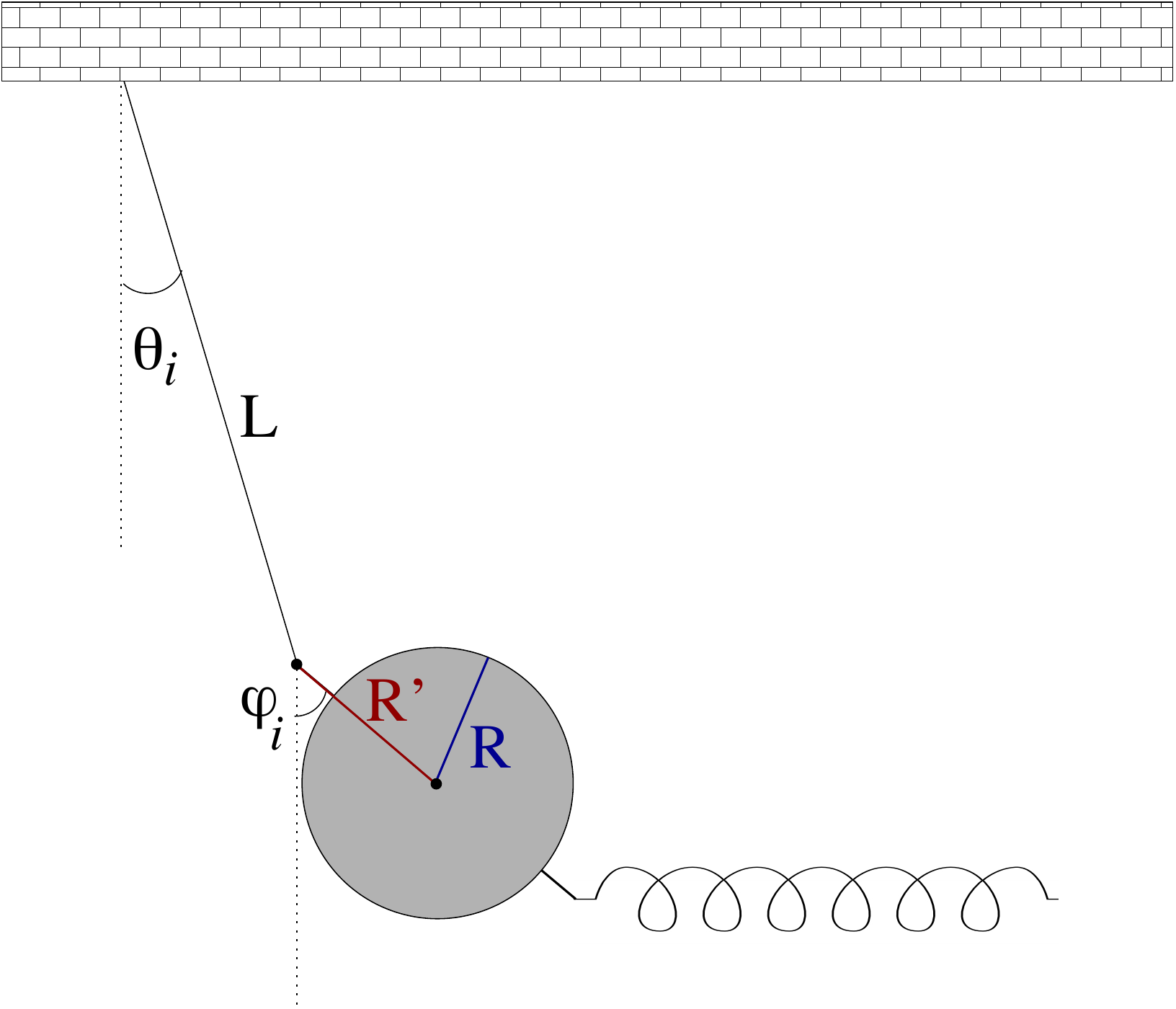}\hspace{1cm}
\includegraphics[width=0.35\textwidth]{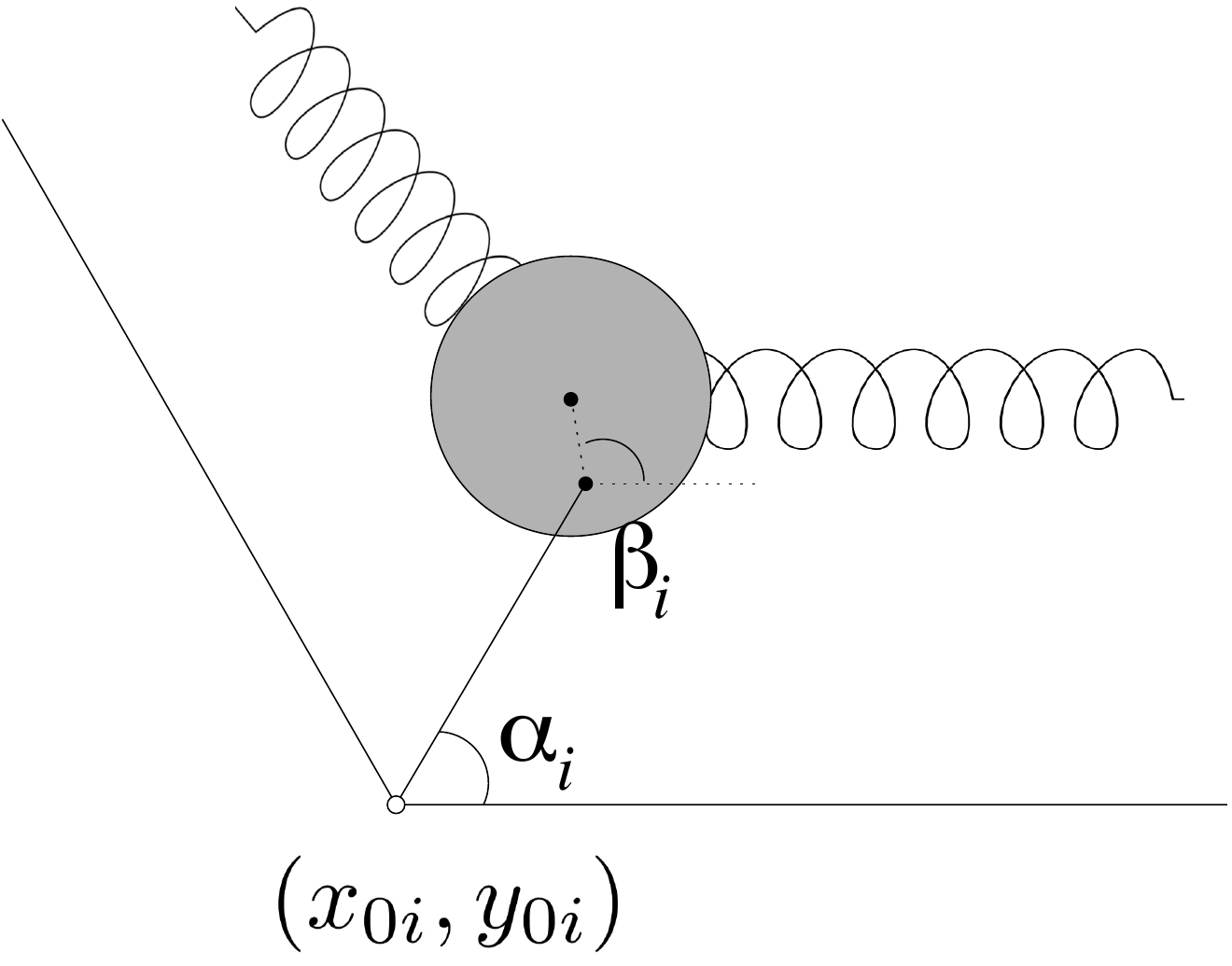}
\caption{Sketch of the coordinates used for representing the motion of the pendula. On the left, a lateral view of the $i-$th pendulum. 
$\theta_i$ is the angle formed by the string with the vertical direction.
The distance along the axis between the centre of mass of the sphere and the center of the hook connecting it to the string is $R'$, which is greater than the radius $R$.
$\varphi_i$ is the angle between the axis of the sphere and the vertical direction.
On the right, a top view of the $i-$th pendulum. 
The suspension point of the pendulum is identified by the coordinates $\left(x_{0i}, y_{0i}\right)$. 
The azimuthal angles $\alpha_i$ and $\beta_i$ are also indicated.}
\label{fig:Lagrangian}
\end{figure}

The sphere can also freely rotate around the top hook, meaning that the axis of the sphere makes a non-zero angle with the direction defined by string of the pendulum, as visible in the left part of \fref{fig:Lagrangian}.
Also visible in \fref{fig:Lagrangian} is the radius $R'$, defined as the distance between the centre of mass of the sphere and the top hook.
Such a distance is larger than the radius of the sphere $R$, because it includes the hook itself.
Since the two hooks have the same length, $R'$ also indicates the distance between the sphere's centre of mass and the bottom hook where the springs are attached.
The experimental value of this distance is $R' = \left(2.85\pm 0.05\right)$~cm.

The $i$-th pendulum is then represented by four coordinates: $\theta_i,\,\varphi_i,\,\alpha_i,\,\beta_i$, as schematically shown in \fref{fig:Lagrangian}.
Two coordinates, $\theta_i$ and $\alpha_i$, are used for defining the position of the point where the string is attached to the sphere.
The other two, $\varphi_i$ and $\beta_i$, define the position of the centre of mass of the sphere. 
Moreover, $\theta_i$ measures the angle between the string of the pendulum and the vertical direction, while $\varphi_i$ measures the angle between the axis of the sphere and the vertical direction, as visible in the left part of \fref{fig:Lagrangian}.
$\alpha_i$ measures the angle between the projection of the string on the $x-y$ plane and the $x-$axis, while $\beta_i$ is measured between the projection of the axis of the sphere on the $x-y$ plane and the $x-$axis.
The two azimuthal angles $\alpha_i$ and $\beta_i$ are indicated in the right part of \fref{fig:Lagrangian}.

\label{appendix2}

We can then express the position of the centre of mass of the sphere with these spherical coordinates:
\begin{equation}
\begin{split}
x^\text{cm}_i &= x_{0i} + L \sin \theta_i \cos \alpha_i +R' \sin \varphi_i \cos\beta_i\\
y^\text{cm}_i &= y_{0i} + L \sin \theta_i \sin \alpha_i +R' \sin \varphi_i \sin\beta_i\\
z^\text{cm}_i &= z_{0i} + L \cos \theta_i +R' \cos \varphi_i.
\end{split}
\end{equation}

It is then straightforward to write the kinetic energy of the $i$-th pendulum as:
\begin{equation}
K_i = \frac{m}{2}\left( \left.\dot{x}^\text{cm}_i\right.^2+ \left.\dot{y}^\text{cm}_i\right.^2 + \left.\dot{z}^\text{cm}_i\right.^2  +\frac{2}{5}R^2 \dot{\varphi}_i^2+ \frac{2}{5}R^2 (\sin\varphi_i)^2  \dot{\beta}_i^2\right),
\label{kinetic_energy_}
\end{equation}
where we have considered the rigid body rotation of the sphere around the joint with the string. Thanks to the bottom position of the hook connecting the sphere to the springs we do not need to include the rotation of the sphere around its axis, as this rotation is decoupled from the motion of the other pendula.

The potential energy of the $i-$th pendulum is defined as $ U^\text{pot}_{i} = - m g z^\text{cm}_i$.
To write down the expression for the elastic potential energy, we have instead to consider the position of the bottom of the sphere, where the springs are attached:
\begin{equation}
\begin{split}
x^\text{s}_i &= x_{0i} + L \sin \theta_i \cos \alpha_i +2R' \sin \varphi_i \cos\beta_i\\
y^\text{s}_i &= y_{0i} + L \sin \theta_i \sin \alpha_i +2R' \sin \varphi_i \sin\beta_i\\
z^\text{s}_i &= z_{0i} + L \cos \theta_i +2R' \cos \varphi_i.
\end{split}
\end{equation}
With these coordinates, the elastic potential energy of the spring that connects the $i-$th pendulum with its nearest-neighbour $j-$th pendulum is:
\begin{equation}
U^\text{s}_{ij} = \frac{k}{2} \left(\sqrt{ \left(x^\text{s}_j -x^\text{s}_i\right)^2+ \left(y^\text{s}_j -y^\text{s}_i\right)^2+ \left(z^\text{s}_j -z^\text{s}_i\right)^2}-\ell_0\right)^2.
\label{potential_energy_}
\end{equation}
We then write down $24$ Euler--Lagrange equations starting from the Lagrangian: $\mathcal{L} = \sum_i \left(K_i - U^\text{pot}_{i} - \sum_{<j>} U^\text{s}_{ij}/2\right)$, where the sum over $j$ is done for the two neighbouring pendula of the $i-$th pendulum.
We notice that this Lagrangian is used as it is and has not been linearised.

By numerically solving the Euler--Lagrange equations with the experimental parameters we obtain the dynamics of the pendula system. 
We then Fourier transform the solutions to obtain numerical frequency spectra.
We observe two sets of frequency peaks. 
The first set is located at low-frequency ($6-16$~rad/s), while the second set is at higher frequency ($35-50$~rad/s).
The low-frequency peaks of the numerical spectra are used in \fref{fig:SOC} to draw the numerical dashed lines. 
As expected, they correspond to modes where the motion of the whole sphere follows that of its upper hook.
On the other hand, the high-frequency modes correspond to oscillations where the spheres have a significant rotation around the top hook, so that the motion of the center of mass is somehow out of phase with that of the top hook. 
In our set-up these latter modes have a lower spectral height and are more damped than the low-frequency modes. 
Although in \fref{fig:exp_spectra_conf} we focus on the low-frequency part of the spectra, both sets of peaks are however visible in the experimental spectra and are found in good agreement with the numerical results.

\end{document}